\documentclass{aa}
\usepackage{graphicx}      
\usepackage{txfonts}
\usepackage{natbib}
\usepackage{epsfig}
\usepackage{subfigure}
\bibpunct{(}{)}{;}{a}{}{,} 

\begin{document}
   \title{[FeII] as a tracer of Supernova Rate in Near-by Starburst Galaxies}

   \author{M.~J.~F.~Rosenberg
          \inst{1}
          \and
          P.~P. van der Werf \inst{1}
          \and
        F.~P. Israel \inst{1}
          }

   \institute{Sterrewacht Leiden, Universiteit Leiden,
              P.O. Box 9513, NL-2300 RA Leiden\\
              \email{rosenberg@strw.leidenuniv.nl}
             }

   \date{Received January 04, 2012; accepted February 16, 2012}

 
  \abstract
   {Supernovae play an integral role in the feedback of
  processed material into the ISM of galaxies and are responsible for
  much of the chemical enrichment of the universe. The rate of
  supernovae can also reveal the star formation histories. Supernova
  rates are usually measured through the non-thermal radio continuum
  luminosity; however, a correlation between near-infrared [FeII] emission and
  supernova remnants has also been noted.}{We aim to find a
  quantitative relationship between the [FeII] at 1.26 $\mu$m ([FeII]$_{1.26}$) luminosity and
  supernova rate in a sample of 11 near-by starburst galaxy
  centers.}{We perform a pixel-pixel analysis of this correlation on
  SINFONI data cubes. Using Br$\gamma$ equivalent width and luminosity
  as the only observational inputs into the Starburst 99 model, we derive the
  supernova rate at each pixel and thus create maps of supernova rates.
  We then compare these morphologically and quantitatively
  to the [FeII]$_{1.26}$ luminosity.}{We have found that a strong linear and
  morphological correlation exists between supernova rate and
  [FeII]$_{1.26}$ on a pixel-pixel basis: \[ log\frac{\nu_{SNrate}}{yr^{-1}pc^{-2}} = 1.01 \pm 0.2\ast  log\frac{[FeII]_{1.26}}{erg s^{-1}pc^{-2}} - 41.17 \pm 0.9\] This relation is valid for normal star forming galaxies but breaks down for extreme ultra luminous galaxies.}{The supernova rates derived from the Starburst 99 model are in good
  agreement with the radio-derived supernova rates, which underlines
  the strength of using [FeII] emission as a tracer of supernova
  rate. With the strong correlation found in this sample of galaxies,
  we conclude that [FeII]$_{1.26}$ emission can be generally used to
  derive accurate supernova rates on either a pixel-pixel or
  integrated galactic basis.}
  
\keywords{Supernova Remnants, Galaxies:nuclei, Near-Infrared} 
\titlerunning{[FeII] as a tracer of Supernova Rate}
\authorrunning{Rosenberg, M.~J.~F. et al.}

   \maketitle
%

\section{Introduction}

Supernova rates (SNrates) are typically estimated by the integrated
non-thermal radio continuum emission from their remnants
(\citealt{1992ARA&A..30..575C}, and references therein).  The
connection between radio continuum emission and SNrate comes from the
tight infrared-radio relation that is usually interpreted to be a
natural consequence of (massive) star formation and stellar evolution
(\citealt{1992ARA&A..30..575C}, and references therein).  Observations
of near-by starburst galaxies show many compact radio sources,
generally attributed to supernova remnants (SNR).
\citet{1990ApJ...357...97C} use observations of SNR in our own galaxy
to establish a relation between the non-thermal radio continuum and SNrate.
\citet{1994ApJ...424..114H} find a similar relation using
observations of compact radio sources in M82.

In addition, near-infrared (NIR) observations of SNR often show strong
[FeII] emission line flux at 1.257 and 1.644 $\mu$m coincident with
the radio peak \citep{1987ApJ...313..847G,1990ApJ...352..172G,1989A&A...214..307O,1990A&A...240..453O}.
In interstellar space, iron atoms are typically locked in dust grains.
However, shock fronts associated with SNRs cause efficient grain
destruction through thermal sputtering.  This releases the iron into
the gas-phase where it is singly ionised by the interstellar radiation
field.  In the extended post-shock region, [FeII] is excited by
electron collisions \citep{2000ApJ...528..186M}, making it a strong
diagnostic line for tracing shocks.  Since [FeII] is commonly
associated with SNRs, the [FeII]/Br$\gamma$ line intensity ratio is
often used to estimate the abundance of SNRs relative to current star
formation.  \citet{1987ASSL..134..283S}'s shock models of [FeII]/H$\beta$
imply a [FeII]/Br$\gamma$ ratio of 20-70 \citep{1988A&A...203..278M},
which is in good agreement with SNR observations
\citep{1987ApJ...313..847G,1988srim.conf..391M}.  Since the [FeII]
emission lifetime of a SNR is approximately $10^4$ years
\citep{1989A&A...214..307O}, the [FeII] luminosity should also be a
strong tracer of the SNrate.  Qualitatively, the connection between
[FeII] session and supernova activity is supported by a wealth of
galactic and extragalactic NIR spectroscopy
\citep{1988srim.conf..391M,1991ApJ...383..164G,1997ApJ...479..694V}
and high-resolution imaging \citep{1993ApJ...406L..11F,
  1997ApJ...485..438G,2000ApJ...532..845A}, which both demonstrate
that [FeII] emission often coincides with known radio SNRs and may
indeed provide at least a relative measure of SNrate.

However, the quantitative connection between NIR [FeII] emission and
supernovae has proved to be more complex and a better understanding of
the underlying physics is necessary.  Radio continuum and [FeII]
emission do not always correspond, even in a specific SNR.  For
instance, \citet{1997ApJ...485..438G} found that there is little
correlation between the radio continuum and [FeII] line emission in
radio-loud SNRs in M82 and suggest that the SNR emission
characteristics change as a function of age.  In addition,
\citet{2003AJ....125.1210A} obtained high-resolution HST [FeII] images
of M82 and NGC 253 and found compact [FeII] emission in only 30-50\%
of radio all SNRs.  They noted that up to 73-78\% of the [FeII] flux
is actually diffuse emission, and does not originate in the discrete
radio SNR which makes it more difficult to interpret the connection
between SNR and [FeII] emission. Vermaas \& van der Werf (in
preparation) argue that the absence of [PII] line emission at 1.19
$\mu$m implies that [FeII] emission is the outcome of grain
destruction by strong shocks.  This diagnostic allows the
determination of the excitation mechanism of [FeII] Vermaas $\&$ van
der Werf (in preparation) find in M83 that although the [FeII] is
diffuse, the distinct lack of [PII] indicates that this emission
nevertheless is the consequence of grain destruction by evolved SNRs
which have lost their individual identities.

\citet{1993ApJ...405..522V}, \citet{1997AJ....113..162C} and
\citet{1997ApJ...479..694V} 
have attempted to correlate SNrates to [FeII] emission by finding a
relationship between the non-thermal radio continuum and [FeII] flux
and using a conversion factor to derive a SNrate.
\citet{1993ApJ...405..522V} uses 6 cm radio fluxes of NGC 6240 from
\citet{1990ApJ...365..478E} to estimate SNrates with the conversion
factor from \citet{1990ApJ...357...97C} which was based on galactic
SNR observations.  \citet{1997AJ....113..162C} adapted the relation
from \citet{1993ApJ...405..522V} to [FeII]$_{1.26}$.
\citet{1997ApJ...479..694V} demonstrated that in blue dwarf galaxies,
all of the [FeII] emission is plausibly accounted for by SNRs and
presented a relationship between [FeII] luminosity and SNrates based on
the M82 calibration.  They also found a relationship between [FeII] and
age, assuming that most of the [FeII] derives from SNRs.

Thus far, there has been little agreement between the radio continuum
and [FeII] line emission in terms of determining SNrates.  However,
such a relationship is desirable because it allows us to use
relatively straightforward NIR [FeII] observations to estimate
SNrates whenever individual radio SNRs cannot be resolved.  In this
paper, we will show that a well-determined relationship between [FeII]
and SNrate exists, and explore and constrain it in a sample of eleven
central regions of near-by starburst galaxies.

First, we will briefly introduce the galaxies in our sample in
Section~\ref{sec:sample} and outline the observational parameters in
Section~\ref{sec:obs}.  The results are discussed in
Section~\ref{sec:results}, which also displays the spectra,
Br$\gamma$, FeII, and H$_2$ line-maps and continuum maps.  In
Section~\ref{sec:methods} we explain the methods used in the analysis,
and in Section~\ref{sec:analysis} we present our analysis of the
[FeII]-SNrate connection. Section~\ref{sec:cont} explores the
dependency of our results on the star formation history of the galaxy
and Section~\ref{sec:radio} compares the SNrates derived from the
Starburst 99 model to the more classical radio-continuum-derived
SNrates.  Finally, our conclusions are highlighted in
Section~\ref{sec:conclusion}.

\section{Sample}
\label{sec:sample}
We have selected eleven bright near-by starburst galaxies at distances
ranging from ~10-100 Mpc (see Table~\ref{table:galinfo}).  All are
spiral galaxies with types ranging from S0a to Sd, with the exception
of NGC 520, which is a merging galaxy.  Two of the galaxies in our
sample are seen edge-on, NGC 520 and NGC 3628.  Far-infrared
luminosities range from $1.4 \times10^{10} L_\odot$ to
$3.2\times10^{12} L_\odot$, where Arp 220 with $L>10^{12} L_{\odot}$
must be classified as an Ultra Luminous Infrared Galaxy (ULIRG), and
NGC 6240, NGC 1614 and NGC 7552, all with $L>10^{11} L_{\odot}$ are Luminous Infrared Galaxies (LIRGs).
Our sample does not contain galaxies with a dominant active nucleus, but
it does have four Low-Ionisation Nuclear Emission-line Region
(LINER) galaxies and one weak Seyfert 2 (Sy2) nucleus.

\begin{table*}
 \caption{Summary of galaxy parameters.}
 \centering
\begin{tabular}{|l|c|c|c|c|c|c|c|c|}
\hline
Galaxy   & RA\tablefootmark{a} & DEC \tablefootmark{a}& Morph.\tablefootmark{b} & Activity \tablefootmark{c}&Dist.\tablefootmark{d} & \textit{cz}\tablefootmark{e} & \textit{i}\tablefootmark{f} &log L$_{FIR}$ \tablefootmark{g}\\
        & J2000 & J2000 & J2000 & & [Mpc] & [km s$^{-1}$] & [$^\circ$] & [L$_\odot$]\\
\hline
\hline
NGC 3628 &11h20m17.0s&+13d35m23s&Sb pec edge-on&HII LINER&12.8&847$\pm$2&79.29&10.14\\
NGC 4536 &12h34m27.0s&+02d11m17s&SAB(rs)bc&HII Sbrst&15.4 &1802$\pm$3&58.9&10.17\\
NGC 1792 &05h05m14.4s &-37d58m51s&SA(rs)bc&HII&13.2&1210$\pm$5&62.78&10.22\\ 
NGC 1084 &02h45m59.9s&-07d34m42s&SA(s)c&HII&16.6&1409$\pm$4&46&10.42\\ 
NGC 1808 &05h07m42.3s&-37d30m47s&(R)SAB(s)a&HII&12.3&1001$\pm$4&83.87&10.55\\ 
NGC 520  &01h24m35.1s&+03d47m33s&pec&Merger Sbrst&30.5&2162$\pm$4&77.49&10.81\\
NGC 7552 &23h16m10.8s&-42d35m05s&(R')SB(s)ab&HII LINER&22.5&1611$\pm$6&23.65&11.03\\
NGC 7632 &23h22m00.9s&-42d28m50s&(R')SB(s)0$^0$&HII &19.3&1535$\pm$15&82.44&11.43\tablefootmark{h}\\
NGC 1614 &04h33m59.8s&-08d34m44s&SB(s)c pec&Sbrst&64.2&4778$\pm$6&41.79&11.43\\
NGC 6240 &16h52m58.9s&+02d24m03s&S0-a&LINER&108.8&7242$\pm$45 &73.0&11.73\\
Arp 220  &15h34m57.1s&+23d30m11s&Sd& LINER Sy2&82.9&  5420$\pm$6& 57&12.50\\     
\hline
\end{tabular}
\tablefoot{ \\
 \tablefoottext{a}{Coordinates of galactic nucleus from NED.} \\
 \tablefoottext{b}{NED} \\
 \tablefoottext{c}{\citet{morph}} \\
 \tablefoottext{d}{NED Hubble Flow Distance (Virgo + GA + Shapley where $H_0=73.0\pm5$ km/sec/Mpc) for NGC 520, NGC 7552, NGC 1614, NGC 6240 and Arp 220, metric distances for the rest. }\\
 \tablefoottext{e}{Heliocentric radial velocity (cz) from radio measurement from \citet{LEDA} except
for NGC 7632 which is from optical measurements.} \\
 \tablefoottext{f}{Inclination measured from \citet{LEDA}}\\
\tablefoottext{g}{L$_{FIR}$ from \citet{2003AJ....126.1607S}} \\
\tablefoottext{h}{L$_{IR}$ for NGC 7632 from \citet{2010ApJ...709..884Y}}\\}
\label{table:galinfo}
\end{table*}

\subsection{NGC 3628}
NGC 3628 is a near-by ($\sim 12.8$ Mpc) HII LINER galaxy.  It is of
morphological type Sb pec and is seen edge-on. Although this galaxy's
nucleus is dominated by a central starburst, it may also be host to a
small AGN \citep{2006A&A...460...45G,2001MNRAS.324..737R}. NGC 3628 is
a member of the Leo triplet, a group of 3 closely interacting
galaxies.  Observations of the HI disk reveal a very disturbed
distribution, pointing to the interaction between NGC 3628 and its
near-by companion NGC 3627 \citep{1978AJ.....83..219R,
  1979ApJ...229...83H}.  Radio observations show an extended radio
core of $350\times60$ pc with a predominately non-thermal spectrum
\citep{1980A&AS...41..151H, 1982ApJ...252..102C, 1991A&A...248...12R}.
Its inner kilo-parsec is dominated by a significant concentration of
molecular gas \citep{2009A&A...506..689I}.

\subsection{NGC 4536}
NGC 4536 is a late-type HII/Starburst galaxy that shows [Ne V] line
emission, which may indicate that the nucleus harbours a weak AGN
\citep{2005AJ....130...73H}.  Previous observations indicate vigorous
star formation in the central nuclear region, shown through Br$\gamma$
emission \citep{1988MNRAS.234P..29P}.  The diffuse radio emission with
three central peaks \citep{1990MNRAS.242..379V,2006AJ....131..701L}
may indicate that this occurs in a ring around the nucleus.  There are
two ultra-luminous X-ray sources, one of which is located in the core
of the galaxy and possibly associated with the weak AGN
\citep{2005ApJS..157...59L}.

\subsection{NGC 1792}
NGC 1792 is in a strongly interacting galaxy group with the dominant galaxy
NGC 1808.  NGC 1792 has surprisingly high luminosities, comparable to
those of its more massive partner NGC 1808, at many different
wavelengths \citep{1994ApJ...432..590D}.  Upon closer inspection,
\citet{1994ApJ...432..590D} found a strongly asymmetric star formation
distribution caused by the external trigger of gravitational
interactions with NGC 1808.

\subsection{NGC 1084}
NGC 1084 is a HII driven Sa(s)c galaxy located $\sim 16.6 $ Mpc
away.  More recently, this galaxy has been classified as an Sbc and Sb
based on its B and H band images, respectively
\citep{2002ApJS..143...73E}.  The H band image reveals a bright
nucleus with an elongated bulge.  Many weakly defined spiral arms were
detected along with many bright knots.  NGC 1084 was mapped at 1.49
GHz by \citet{1987ApJS...65..485C}, who detected a strong continuum
source to the south of the nucleus.  \citet{2007MNRAS.381..511R}
derived a star formation rate of 2.8 M$_{\sun}$ yr$^{-1}$ along with
chaotic star formation that is not necessarily confined to the spiral
arms.  The northern and southern regions of the galaxy have different
star formation rates and typical ages.  The north is characterised by
a series of short bursts happening intermittently over the past 40
Myr, possibly due to an interaction with a gas rich galaxy
\citep{2007MNRAS.381..511R}.  The southern half of the galaxy is home
to much younger stars, $< 4 $ Myr \citep{2007MNRAS.381..511R}.

\subsection{NGC 1808}

NGC 1808 is the most massive member of a small group of galaxies.  The
galactic center has a super-wind, compact radio sources, and a
molecular ring
\citep{1996A&A...313..771K}. \citet{1985A&A...145..425V} first
suspected that NGC 1808 has a faint Seyfert nucleus.  This has since
been refuted by many
\citep{1992MNRAS.259..293F,1993AJ....105..486P,1994ApJ...425...72K},
yet there is recent hard- X-ray evidence to support the Seyfert nucleus
\citep{Awaki1993221}.

\subsection{NGC 520}
Using long slit optical photometry and near-infrared imaging,
\citet{1990ApJ...355...59S} uncovered two separate nuclei in NGC 520,
establishing this as a galaxy in the process of merging.
\citet{1991ApJ...370..118S} went on to determine that NGC 520 is most
likely the result of a galaxy collision of a gas-rich and a gas-poor
galaxy that happened $\sim3\times10^8$ years ago.  A thick dust lane
at PA=95$^\circ$ completely obscures the primary nucleus at optical
wavelengths, but the secondary nucleus is visible to the north-west of
the primary.  There are strong plumes of ionised gas, thought to
represent a bipolar outflow away from the starburst-dominated nucleus
\citep{1996ApJ...472...73N}.

\subsection{NGC 7552}
NGC 7552 is a LIRG HII galaxy and host to a weak LINER.  This galaxy
has a starburst ring surrounding the nucleus. It does not appear to be
overly disturbed, which suggests a slow bar-mediated star formation
evolution.  Due to the quiescent nature of the nucleus, NGC 7552
provides a near-perfect environment to study the bar and starburst
ring around it.  In addition to the bar, there are two dominant spiral
arms and two weaker rings at radii of 1.9 and 3.4 kpc
\citep{1990A&A...239...90F}. \citet{1994AJ....107..984F} found an
inner ring with a radius of 1 kpc in the radio, which is not visible
in the NIR continuum but is visible in NIR color maps and Br$\gamma$
emission.  The dominant circumnuclear ring is at a distance of 850 pc
from the center of the nucleus and the spiral arms provide a flow of
molecular gas into this region \citep{1997ApJ...488..174S}.

\subsection{NGC 7632}
NGC 7632 is a member of a loose galaxy group, containing multiple
galaxies.  This is a ring galaxy, which is observed to be distorted
and bent towards the other group members \citep{1981ApJS...46...75A}.
The interacting galaxies located south of the NGC 7632 also appear
strongly distorted with much absorbing material
\citep{1981ApJS...46...75A}.

\subsection{NGC 1614}
NGC 1614 is a SB(s)c peculiar LIRG with both LINER and starburst
activity. Although only one NIR peak has been observed,
\citet{1990AJ.....99.1088N} considered the tidal tails or plumes to be
evidence that NGC 1614 is the result of a merger of at least two
galaxies.  A central nucleus of 45 pc is surrounded by a 600 pc
diameter ring of current star formation
\citep{2001ApJ...546..952A}. The Br$\gamma$ emission reveals a double
peaked morphology \citep{2001A&A...366..439K}, in agreement with the
radio continuum \citep{1982ApJ...252..102C}.  Further studies of NGC
1614 by \citet{2010A&A...513A..11O} found the molecular gas
distribution double peaked at R = 300 pc with an additional weaker
peak in the center of the nucleus.  A radio continuum ring is also
found at R = 300 pc which is triple peaked, each consistent with the
brightness temperature of SNR.  \citet{2010A&A...513A..11O} concludes
that the LINER spectrum is due to shocks associated with supernovae
and therefore most likely caused by star formation and not AGN
activity.

\subsection{NGC 6240}
NGC 6240 just falls short of being a traditional ULIRG with a
luminosity of $L_{IR}=10^{12}L_{\odot}$ but otherwise has all the
characteristics of the class.  The extended tidal tails seen in the
optical indicate a merger.  NGC 6240 hosts a double nucleus separated
by 1.5'' $\pm$ 0.1'' as seen in ground based observations in the
optical \citep{1993A&A...277..416S} and near infrared
\citep{1994ApJ...437L..23D}, yet a separation of 2'' $\pm$ 0.15'' as
measured in [FeII]$_{1.64}$ $\mu$m \citep{1993ApJ...405..522V} and
radio observations
\citep{1990ApJ...362..434C,1990ApJ...365..478E,2001MNRAS.325..151B}.
The high infrared luminosity is partially due to the powerful nuclear
starburst and partially to an AGN continuum
\citep{1998ApJ...498..579G}.

\subsection{Arp 220}
The ULIRG Arp 220 emits 99\% of its bolometric luminosity in the
infrared \citep{1984ApJ...283L...1S,1984Natur.311..237E}.  Arp 220 is
a very gas-rich galaxy, with the highest-density component in the
nucleus, with a molecular gas density up to $\sim10^{-8} cm^{-3}$
\citep{2011ApJ...743...94R}.  This high IR luminosity and intense
starburst activity points to a merger, supported by the existence of
two nuclei, separated by $\sim 1''$ \citep{1997ApJ...484..702S}.  Arp
220 has an extremely high extinction, with estimates running from A$_v
\sim 50$ to 1000 \citep{1996A&A...315L.133S,1998ApJ...507..615D}.
\citet{1998ApJ...493L..17S} resolve the radio nucleus of Arp 220 and
find many radio point sources, which they identify as SNRs. They
deduce a star formation rate of 50-100 M$_\odot$ yr$^{-1}$ and a
corresponding SNrate of 1.75-3.5 yr$^{-1}$.

\section{Observations}
\label{sec:obs}
Observations of the sample galaxies were made with the Spectrograph
for INtegral Field Observations in the Near Infrared (SINFONI) at the
VLT.  SINFONI provides spatial and spectral data in the form of data
cubes in J, H, and K bands.  The SINFONI instrument is mounted at the
Cassegrain focus of the Unit Telescope 4 at the Very Large Telescope
(VLT).

We took observations in all three J, H, and K bands using a spatial
pixel scale of 0.25'' corresponding to a field of view of
8''$\times$8'' and a spectral resolution of 2000, 3000 and 4000
respectively, between October, 2006 and March, 2007.  All science
observations were taken in the ABA'-nodding mode (300s of object-300s
of sky-300s of object), where A' is slightly offset from A.  The
object exposures are averaged during the reconstruction of the data
cube.

We extracted additional observations from the SINFONI archive
(http://archive.eso.org/wdb/wdb/eso/sinfoni/form).  These observations
include five galaxies, Arp 220, NGC 1614, NGC 1808, NGC 6240 and NGC
7552.  As selection criteria we required the galaxies to be bright and near-by,
and to have archival observations in the 0.25'' spatial
resolution mode in all three bands and 300s integration times.

We used the standard reduction techniques of the SINFONI pipeline on
all observations. including corrections for flat field, dark current,
non-linearity of pixels, distortion, and wavelength calibration.  We
obtained the flux calibration and atmospheric corrections from
observations of a standard star.  Finally, we calibrated the continuum
fluxes to match the published 2MASS observations in the same aperture,
and we determined a ``flux correction factor'' for each galaxy in each
observed band, and we applied these correction factors to the line
intensities as well.

\section{Results}
\label{sec:results}

\subsection{Spectra}
\begin{figure*}
\includegraphics[width=17cm]{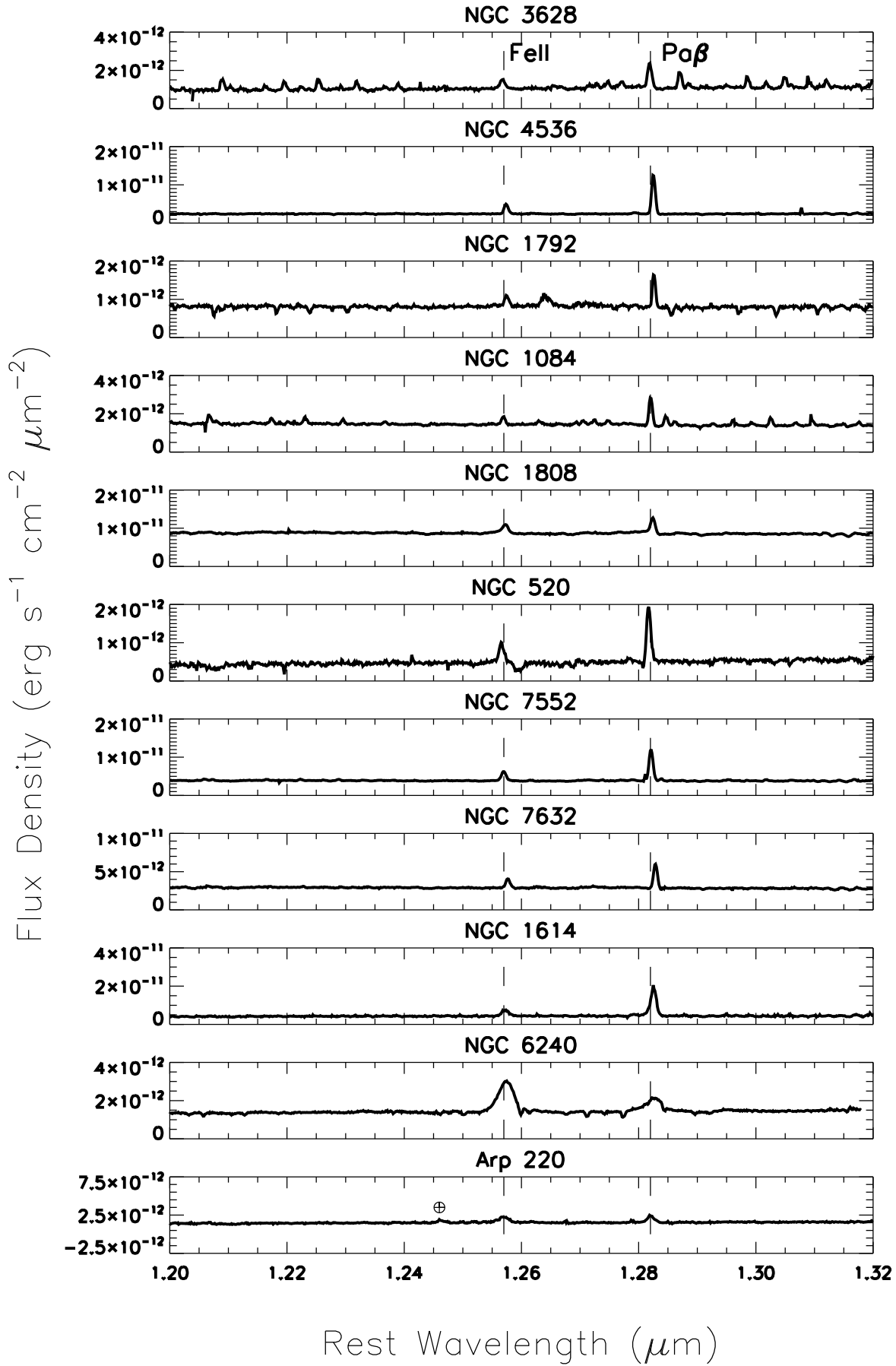}
\caption{Spatially integrated rest-frame spectra from the nucleus of
  each galaxy in the J band.  The spatial area used in the integration is shown in
  Figure~\ref{fig:data1} and \ref{fig:data2} by a black rectangle.  Each spectrum is integrated over an 8x8 pixel area.
  FeII$_{1.26}$ and Pa$\beta$ emission lines are denoted by dashed
  lines and marked along the top of the figure.  The broad feature
  near 1.25 $\mu$m in the Arp 220 spectra is an atmospheric artifact
  along with the narrow absorption features in NGC 520, and the
  numerous broad peaks between 1.2-1.24 and 1.29-1.32 $\mu$m in the NGC
  3628, NGC 1792 and NGC 1084 spectra. }
\label{fig:spectraJ}
\end{figure*}

\begin{figure*}
\includegraphics[width=17cm]{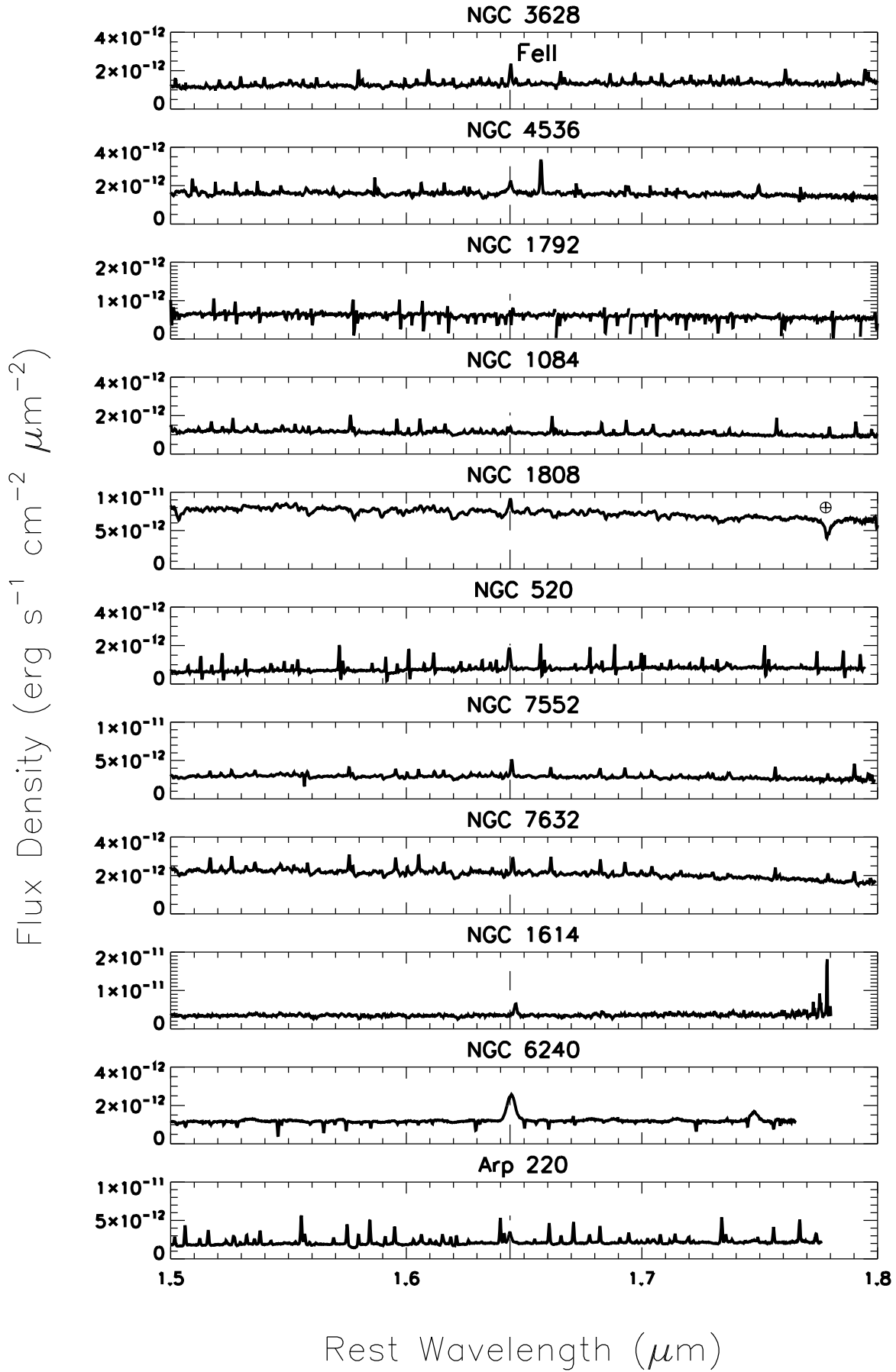}
\caption{Spatially integrated rest-frame spectra from the nucleus of
  each galaxy in the H band.  The spatial area used in the integration is shown in
  Figure~\ref{fig:data1} and \ref{fig:data2} by a black rectangle.  Each spectrum is integrated over an 8x8 pixel area.  The
  FeII$_{1.64}$ emission line is denoted by a dashed line and marked
  along the top of the figure.  Several spectra are degraded by
  residual atmospheric OH features.}
\label{fig:spectraH}
\end{figure*}

\begin{figure*}
\includegraphics[width=17cm]{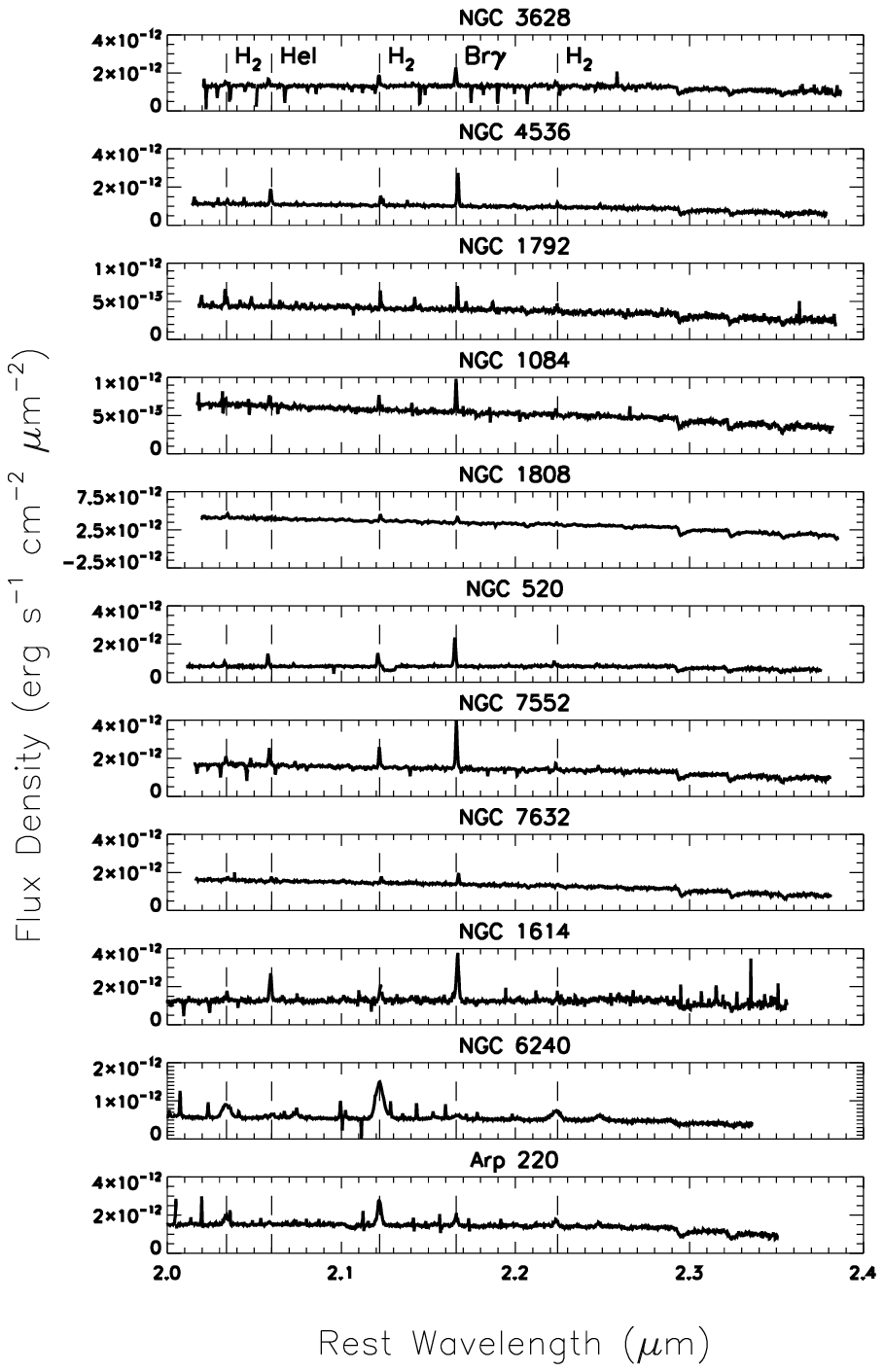}
\caption{Spatially integrated rest-frame spectra from each galaxy in
  the K band.  The spatial area used in the integration is shown in
  Figure~\ref{fig:data1} and \ref{fig:data2} by a black rectangle.  Each spectrum is integrated over an 8x8 pixel area. Br$\gamma$, H$_2$, and
  HeI$_{2.06}$ emission lines are denoted by dashed lines and marked
  along the top of the figure.  Residual atmospheric contamination is
  present in the form of narrow peaks in the spectra of NGC 1614 and
  Arp 220 and as sharp absorption features in the spectra of NGC 3628
  and NGC 7552.}
\label{fig:spectraK}
\end{figure*}

We show the spectra of each galaxy center integrated over a high
signal-to-noise region in the J (Figure~\ref{fig:spectraJ}), H
(Figure~\ref{fig:spectraH}) and K bands (Figure~\ref{fig:spectraK}).
The actual spatial area over which the spectra were integrated is
over-plotted in the K band images in Figure~\ref{fig:data1} and
\ref{fig:data2}.  We have focused on non-nuclear regions to illustrate the spectral signature of the extended emission.  The dominant emission lines in each band are marked
with dotted lines, notably FeII$_{1.26}$ and Pa$\beta$ in the J band,
FeII$_{1.64}$ in the H band and Br$\gamma$, HeI$_{2.06}$, and multiple
H$_2$ lines in the K band.

\subsection{Continuum and line-maps}
The SINFONI data-cubes allow us to construct J, H and K continuum maps
as well as line-maps for each of the emission lines detected in these
bands.  These emission lines provide important diagnostics to trace
specific physical processes.  The Pa$\beta$ (1.282 $\mu$m) and
Br$\gamma$ (2.166 $\mu$m) HI lines trace massive, young star
formation.  The [FeII] emission lines, which emit most strongly at
1.257 and 1.644 $\mu$m, are commonly used as tracers of strong shocks
associated with supernova remnants, nuclear winds, or jets.  There is
also a wealth of ro-vibrational H$_2$ lines throughout the H, and K
bands, the brightest of which occur at rest wavelengths of 2.122
$\mu$m (1-0 S(1)), 2.248 $\mu$m (2-1 S(1)), and 2.034 $\mu$m (1-0
S(2)).  H$_2$ can be excited by UV florescence from massive stars, or
thermally by shocks from supernovae or stellar winds.  The relative
intensities of the various H$_2$ lines indicates which of these
physical processes is exciting the gas.

\begin{figure*}
\includegraphics[width=17cm]{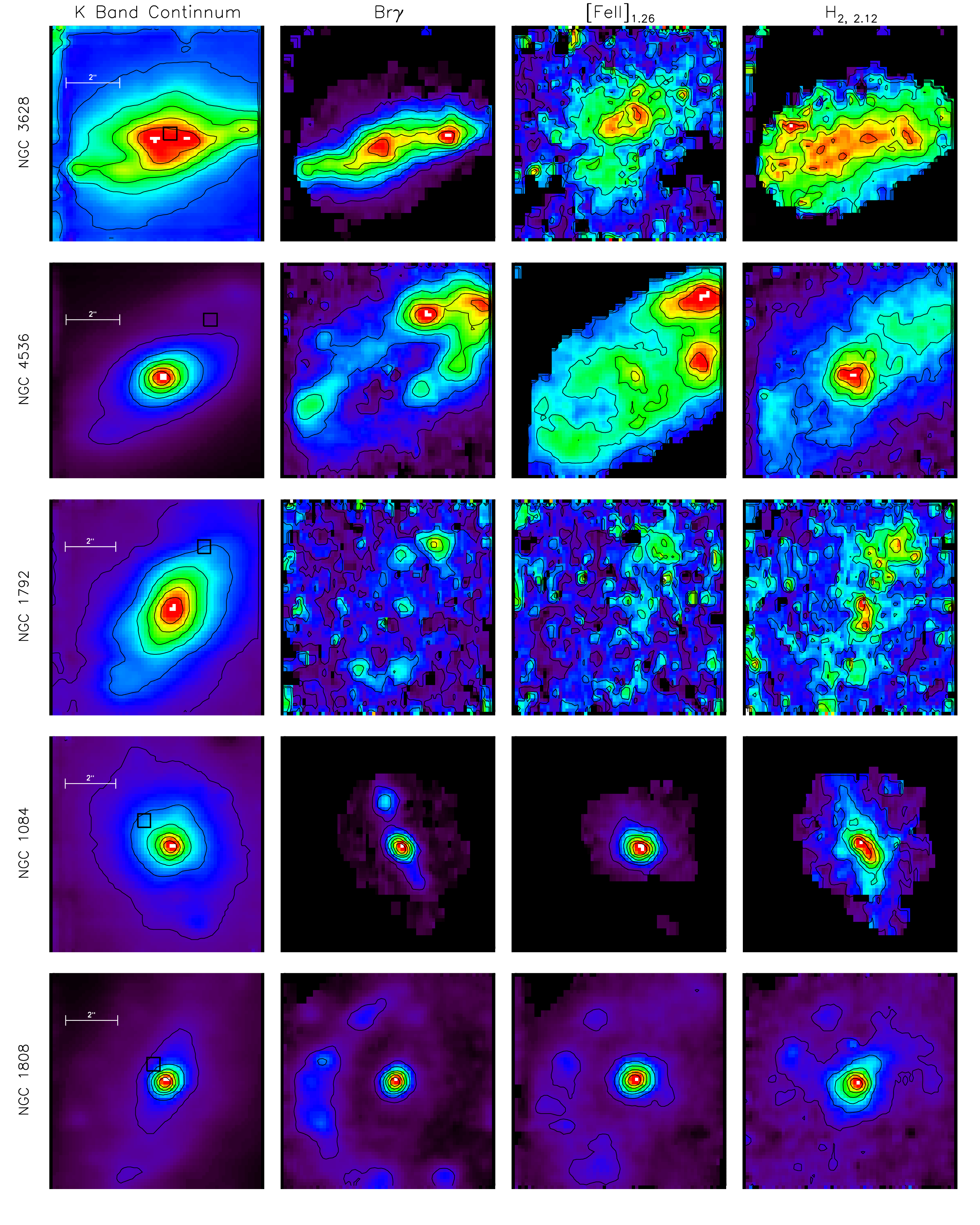}
\caption{K band continuum map, Br$\gamma$, [FeII]$_{1.26}$, and H$_{2,
    2.12}$ line maps of NGC 7632, NGC 3628, NGC 4536, NGC 1792, and NGC
  1084.  The pixel scale is given by the 2'' scale bar in the K band
  column.  The black rectangle marks the areas
  over which the spectra in Figures~\ref{fig:spectraJ}, \ref{fig:spectraH} and \ref{fig:spectraK} were integrated. In
  all figures, north is up.}
\label{fig:data1}
\end{figure*}

\begin{figure*}
\includegraphics[width=17cm]{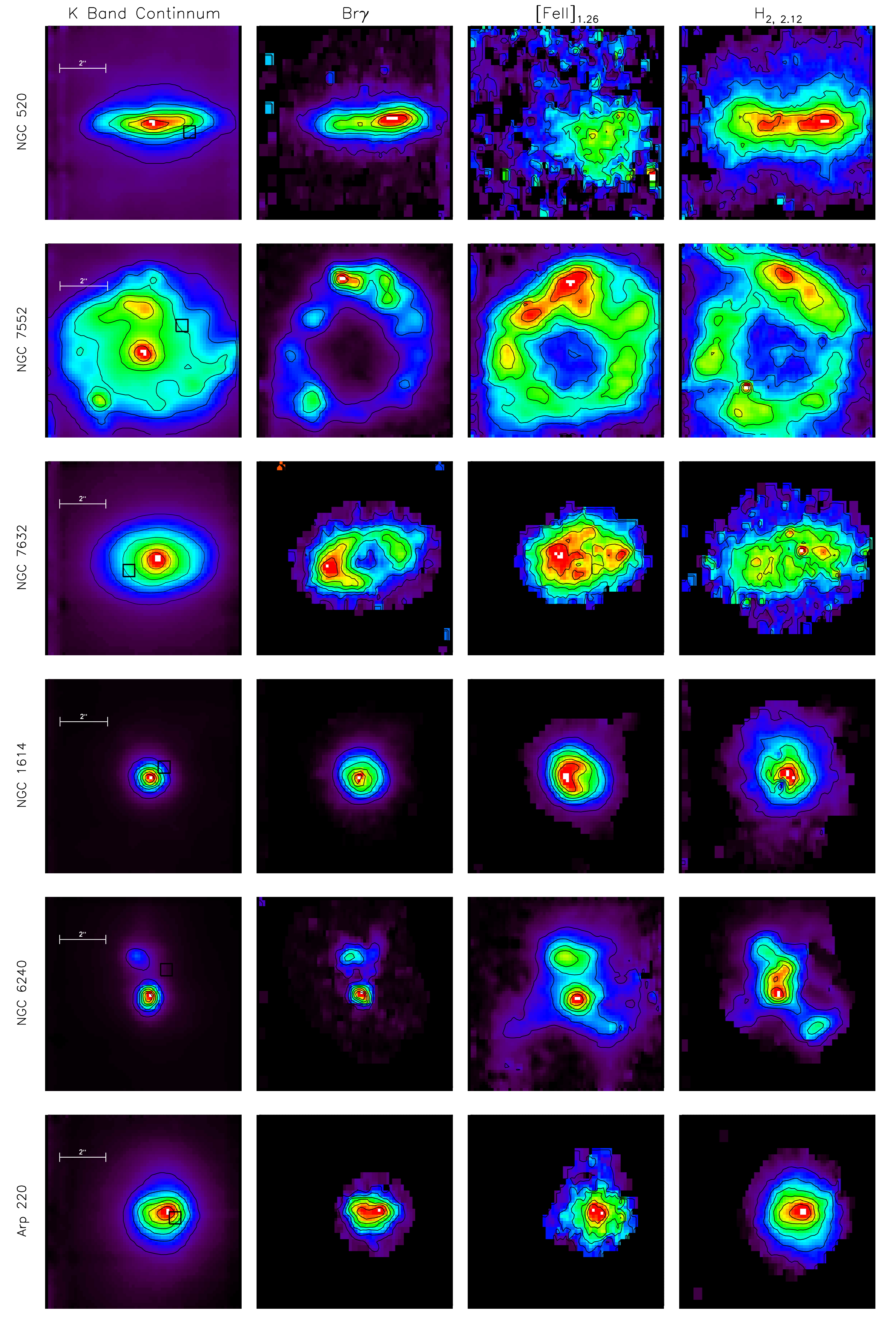}
\caption{K band continuum map, Br$\gamma$, [FeII]$_{1.26}$, and H$_{2,
    2.12}$ line maps of NGC 1808, NGC 520, NGC 7552, NGC
  1614, NGC 6240 and Arp 220.  The pixel scale is given by the 2''
  scale bar in the K band column.  The black
  rectangle represents the area over which the spectra in
  Figures~\ref{fig:spectraJ}, \ref{fig:spectraH} and
  \ref{fig:spectraK} were integrated.  In all figures, north is up.}
\label{fig:data2}
\end{figure*}

In Figure~\ref{fig:data1} and Figure~\ref{fig:data2} we present the K
band continuum, Br$\gamma$, [FeII]$_{1.26}$ and H$_{2, 2.12}$ line
maps for each of the galaxies in our sample.  The continuum maps were
created by finding the average continuum level at each pixel,
excluding any emission lines.  The line-maps were created using
QFitsView and DPUSER's evaluated velocity map function ``evalvelmap'',
developed by the Max Planck Institute for Extraterrestrial Physics and
available at http://www.mpe.mpg.de/~ott/QFitsView/.  This function
fits a gaussian profile to the emission line at each pixel.

The first galaxy in Figure~\ref{fig:data1} is the edge-on NGC 3628,
which has a strong central dust lane running from east to west.  This
is best seen by comparing the Br$\gamma$ and [FeII] maps.  Since
K-band Br$\gamma$ emission is at a longer wavelength, it is less prone
to extinction and the morphology resembles that of the K band
continuum, with an additional emission peak at the western edge of the
galaxy.  The J-band [FeII] emission is at a shorter wavelength and is
more affected by the dust lane.  It shows an asymmetric hour glass
morphology, with the brightest regions above the K band nucleus.  The
H$_2$ emission is similar to the Br$\gamma$ emission, but it has more
diffuse morphology with strong emission throughout the very central
region.

NGC 4536 reveals a star forming ring best seen in Br$\gamma$ and
[FeII] emission.  The H$_2$ emission is concentrated at the K band
continuum peak, with extended diffuse emission throughout the ring.
Both Br$\gamma$ and [FeII] peak in the north-west corner of the
galaxy but their peaks do not coincide.

NGC 1792 has a K-band continuum morphology similar to that of NGC
4536, but the line emission is much weaker in this galaxy.  The low
signal-to-noise ratios prevent us from identifying any real structure.
However, in the Br$\gamma$ map a small peak occurs at the K band
continuum maximum, with structures suggesting spiral arms emanating
from the nucleus.  Although both H$_2$ and [FeII] emission peaks at
the end of these ``arms'', the actual [FeII] fluxes hardly exceed the
noise level.  The H$_2$ has a higher signal-to-noise than the other
emission lines and there are is distinct H$_2$ peaks emission in the
nuclear region.

In NGC 1084, all three emission lines appear to peak on the K-band
continuum maximum.  The Br$\gamma$ and H$_2$ maps show a somewhat
similar distribution, with extended diffuse emission along the major
axis line, and a weak secondary peak to the north.  The [FeII] map
shows a single central maximum surrounded by weak diffuse emission.

NGC 1808 also has a star-formation ring, weakly seen in both the
Br$\gamma$ and [FeII] maps and a few local maxima also seen in the H$_2$
map.  The ring appears to be very asymmetric, with relative bright
fluxes  at the eastern edge and almost no emission from the western
edge.  The H$_2$ emission is distributed more symmetrically around the
K band peak.

In Figure~\ref{fig:data2}, we first examine NGC 520, another nearly
edge-on galaxy that also hosts a dense dust lane running east to west.
The Br$\gamma$ morphology is very asymmetric, peaking to the west of
the K-band continuum maximum.  The H$_2$ emission peaks at the same
location as the Br$\gamma$, but has a secondary peak east of the
K-band center.  The [FeII] emission suffers from extinction by the
dust lane and only weak [FeII] emission is seen to the south-west of
the galaxy mid-plane.

NGC 7552 is a face-on galaxy with a very clear star-formation ring
around the nucleus.  The Br$\gamma$ emission shows a pronounced
minimum in the center, but clear individual maxima in the ring.
Unlike \citet{1994ApJ...433L..13F}, who found the H$_2$ morphology to
match that of the K band continuum, we find the distributions of the
H$_2$ and the Br$\gamma$ emission to be very similar, with an arm or
bar-like structure extending from the northern Br$\gamma$ peaks.  The
[FeII] emission is different: it shows roughly the same ring-like
shape, but with emission peaks in very different positions.

NGC 7632 reveals an asymmetric ring in Br$\gamma$ emission.  The
emission is much stronger on the east side of the galaxy in both
Br$\gamma$ and [FeII] emission. The east side is in the direction of
the other group members (\citealt{1981ApJS...46...75A}).  The
distortion and asymmetry is most likely evidence of the tidal
interaction with its partner galaxies.

NGC 1614 has a very compact and radially symmetric morphology in the K
band, Br$\gamma$ and H$_2$ emission lines.  However, the [FeII]
emission shows a crescent morphology around the nucleus.

The double nucleus of NGC 6240 is evident in the K band continuum and
the Br$\gamma$ line images. Less clearly, the double nucleus also
shows up in the H$_2$ and [FeII] images that reveal a more complex
morphology.  In all images, strongest emission coincides with southern
nucleus.

Arp 220 is another compact galaxy that appears fairly symmetric in the
K band and all the emission line maps.  The peak is slightly off
center towards the western side of the galaxy and there are two
separate peaks in the [FeII] emission.

\section{Methods}
\label{sec:methods}
To further study the correlation between [FeII] emission and SNrate,
we have performed a pixel by pixel analysis.  We evaluated the
continuum flux, emission line strength and equivalent width at each
pixel in the SINFONI data-cubes in order to gain insight into the
variation of physical properties (such as ages and excitation
mechanisms) across each galactic nucleus.  We first corrected the
line-maps for extinction, and used the Br$\gamma$ equivalent widths as
an input into Starburst 99 to calculate SNrates. These could then be
correlated with extinction-corrected [FeII] luminosities.

\subsection{Extinction correction}

In order to study the intrinsic emission line strengths, it is
necessary to determine the degree to which dust obscures each nucleus.
We accomplish this by identifying emission lines with fixed intrinsic
intensity ratios, and comparing the observed to the intrinsic line
ratio.  In our NIR database, the best line pair to use is Pa$\beta$ (H
5-3) and Br$\gamma$ (H 7-4), which has a ratio of 5.88
\citep{1987MNRAS.224..801H} over a wide range of physical conditions.
Extinction is also probed by the [FeII]$_{1.26}$/[FeII]$_{1.64}$ ratio, in
which both [FeII] lines originate from the same upper level with an
intrinsic ratio of 1.36 \citep{1988A&A...193..327N}.  
For completeness sake, we calculated both line ratios and compared the extinction maps.  There is little disparity in morphology between the [FeII] derived and HI dervied extinction maps. 
It is clear, however, that the [FeII] lines are of limited use, as the spectral baseline
they define is small and provides relatively little differential
extinction. More importantly, the [FeII] lines have lower signal-to-noise
ratios than the J and K lines, the [FeII]$_{1.64}$ being particularly bad in
this respect as it is degraded by residual
atmospheric OH contamination in many of our observations.  As a result, meaningful extinction maps based on the [FeII] line ratio could only be derived over limited areas for a minority of the galaixes.  Thus, the
extinctions derived from the Pa$\beta$/Br$\gamma$ ratio should be the
more reliable and will be used to complete the following analysis.

\begin{figure*}
\centering
\includegraphics[width=17cm]{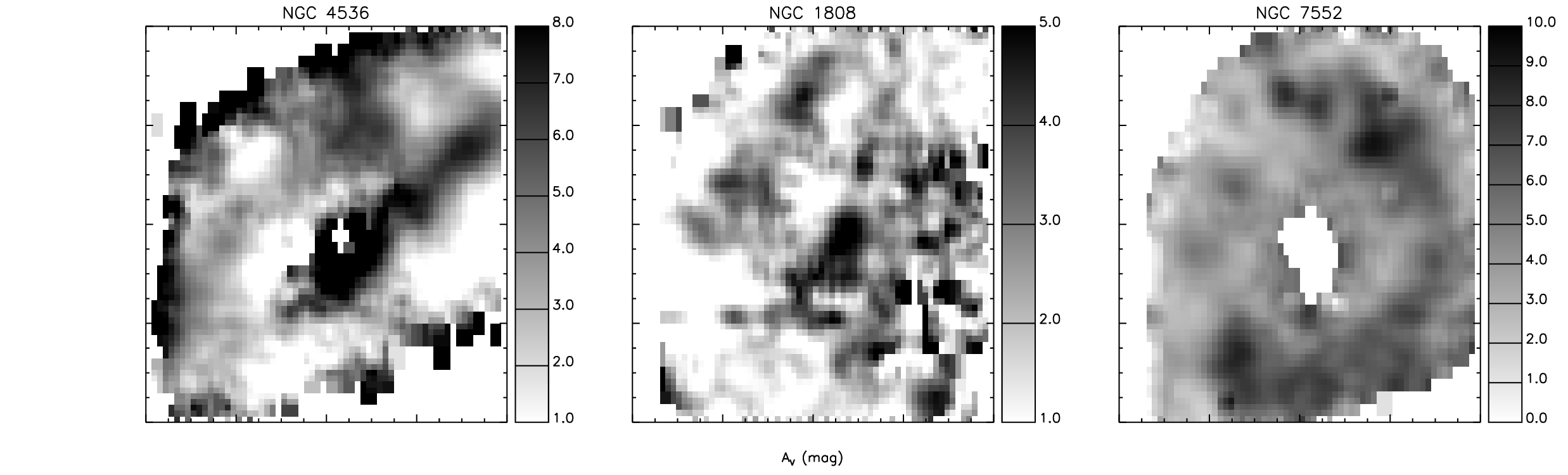}
\caption{Extinction maps of NGC 4536, NGC 1808 and NGC 7552.  The color scales are displayed to the right of each map, the values represent $A_V$ measured in magnitudes.}
\label{fig:ext}
\end{figure*}

We calculated the extinction at each pixel and created an extinction
map revealing the regions most effected by dust. Extinction maps for three example galaxies are shown in Figure~\ref{fig:ext}.
Table~\ref{tab:lineflux} lists the averaged visual extinction (A$_V$)
over the observed field of view for both line ratios.  The Table also
includes the integrated, extinction-corrected Br$\gamma$, Pa$\beta$,
and [FeII] line fluxes.  The maps were integrated over the full field
of view but were filtered for low signal-to-noise pixels and pixels
with line-widths either too narrow or too broad.  The visual extinction
A$_V$ is determined assuming using the near-infrared extinction law
$A_{\lambda}\propto\lambda^{-1.8}$ \citep{1990ApJ...357..113M}.  The
disparity between the extinctions derived from the Fe and H lines
mostly reflects the poor quality of the former, as discussed above. In
the case of NGC 1792, the little line emission observed has a very low
signal-to-noise ratio (cf. Figure~\ref{fig:data1}), negatively
affecting the accuracy of any A$_V$ measurement, hence also the
accuracy of the extinction corrected [FeII] and Br$\gamma$ fluxes.  In
NGC 6240, the extremely low Br$\gamma$ flux makes it difficult to
accurately determine the extinction on a pixel by pixel basis.

\begin{table*}
\caption{Unextinction corrected Br$\gamma$, Pa$\beta$, [FeII]$_{1.26}$, and [FeII]$_{1.64}$ fluxes with the A$_V$ derived by the Pa$\beta$/Br$\gamma$ intrinsic line ratio.  All measurements have a 10\% error, which is the estimated calibration uncertainty.  NGC 1792 has a 20\% error due to the low signal to noise of this observation.}
 \begin{tabular}{|l||c|c|c|c|c|}
\hline
Galaxy&      Pa$\beta$  &    Br$\gamma$   &[FeII]$_{1.26}$     & [FeII]$_{1.64}$   &Av$_{Br\gamma}$ \tablefootmark{a}\\
      & 10$^{-14}$ (erg s$^{-1}$cm$^{-2}$) & 10$^{-14}$ (erg s$^{-1}$ cm$^{-2}$)&10$^{-14}$ (erg s$^{-1}$ cm$^{-2}$)& 10$^{-14}$(erg s$^{-1}$ cm$^{-2}$) & (mag) \\
\hline
NGC 3628& 0.7    &   1.4  &   1.4  &   0.6 &   20.3 \\
NGC 4536& 12.7    &   3.5    &   3.4  &   4.5 &   3.9 \\
NGC 1792& 1.3    &   0.3     &    0.1  &   --  &  2.9 \\
NGC 1084& 1.0    &   0.3      &   0.3  &   0.5&   3.0  \\
NGC 1808& 18.3    &   4.4    &   9.2 &   8.0&   2.7 \\
NGC 520 & 1.2    &   1.3   &   1.2  &   0.3&   15.3  \\
NGC 7552& 20.7    &  6.8     &   6.3 &   8.1 &   5.3 \\
NGC 7632& 1.8    &   0.5      &   0.5  &   0.8 &   3.9\\ 
NGC 1614& 13.8    &   3.5    &   3.1  &   3.7 &   3.3 \\
NGC 6240& 3.3   &   0.6     &   8.5 &   8.7 &    15.6 \\
Arp 220 & 0.4   &   0.5  &    0.7  &   --   &   15.4  \\
\hline
 \end{tabular}
\tablefoot{ \\
 \tablefoottext{a}{A$_V$ is determined assuming using the near-infrared extinction law $A_{\lambda}\propto\lambda^{-1.8}$ \citep{1990ApJ...357..113M}.  Although the measured A$_V$ is very high in some galaxies, the extinction measured in the NIR wavelength range is significantly less.} \\}
\label{tab:lineflux}
\end{table*}

In order to obtain the best possible result, we constructed
extinction-corrected [FeII]$_{1.26}$ and Br$\gamma$ line maps using
pixel-by-pixel extinctions derived from the Pa$\beta$/Br$\gamma$ line
ratio only.  We also constructed pixel-by-pixel maps of the Br$\gamma$
equivalent width directly from the observed Br$\gamma$; these are
therefore independent of the assumed extinction. The Br$\gamma$
equivalent width and the Br$\gamma$ luminosity thus determined are the
\emph{only} observational inputs used in the calculation of SNrates
from the Starburst 99 model, described in detail in
Section~\ref{sec:sb99} below.

\subsection{Calculating SNrate}
\label{sec:sb99}
Starburst 99 (from here on referred to as SB99) is a tool that models
spectrophotometric properties of star-forming galaxies \citep{sb99}
such as spectral energy distributions (SEDs), luminosities, equivalent
widths, supernova rates and colours.  This includes predictions of the
variations in these properties as a function of starburst age.  The
models have been calculated for 5 different metallicities
($Z=0.04,0.02,0.008,0.004,0.001$) and 3 different initial mass
functions (IMF).  In addition, the two extreme star-formation modes
are considered: the {\it continuous} mode in which star formation
proceeds continuously at a constant rate, and the {\it instantaneous}
mode where it has the form of a delta-function starburst.

In our analysis, we assumed a near-solar metallicity ($Z=0.02$), a
Salpeter IMF ($\alpha=2.35, M_{low}=1 M_{\odot}, M_{up}=100
M_{\odot}$) and an instantaneous star formation mode.  In the case of
an instantaneous starburst, the model normalises the burst to an
initial starburst mass of $10^6$ M$_{\odot}$.  We calculated the
average age of the population dominating the emission in each pixel
from the observed Br$\gamma$ equivalent width (EW(Br$\gamma$)), which
is independent of this normalisation. SB99 provides the expected
SNrate as a function of the age thus determined.  Due to the
normalisation of $10^6$ M$_{\odot}$, the SNrate must be appropriately
scaled by comparing the SB99 age-dependent prediction of ionising
photon flux (N(H$^{\circ}$)) to the observed ionising photon flux,
which in turn scales with Br$\gamma$ luminosity.  This comparison
yields a scaling factor that is directly proportional to the initial
mass and initial star formation rate. This allowed us to determine the
actual SNrates based on the true initial conditions of the region
represented in each pixel.

\section{Analysis}
\label{sec:analysis}
In this section, we use the observed line-maps and the SB99 results to
do a detailed analysis of each galaxy in order to better understand
the complex relationship between [FeII] emission and SNrate.  First,
we present the age and SNrate estimates for each galaxy.  Then we will
provide both a qualitative and quantitative comparison of the SNrates
and the [FeII] luminosity.

\subsection{Age and SNrate}

Table~\ref{tab:snrate} lists the Br$\gamma$ equivalent widths as well
as the average age and integrated SNrate using the SB99 instantaneous
burst model for each galaxy.  Although the analysis was done on a
pixel-pixel basis, the values given in Table~\ref{tab:snrate} are
either averaged (in the case of age and equivalent width) or
integrated (in the case of the SNrate) over the galaxy.

\begin{table}
\caption{Average equivalent width of Br$\gamma$ and the SB99 derived average ages and integrated SNrates using the instantaneous starburst model.}
 \begin{tabular}{|l||c|c|c|}
\hline
Galaxy  &  EW(Br$\gamma$)  &    Age$_{inst}$   & SNrate$_{inst}$  \\ 
        &      \AA{}       &   (Myr)        &   yr$^{-1}$     \\
\hline
NGC 7632& 5.9             &   7.0           &    0.03        \\
NGC 3628& 7.5             &   6.9           &    0.01         \\
NGC 4536& 10.6            &   6.7           &    0.3        \\
NGC 1792& 4.1             &   7.9           &    0.007         \\
NGC 1084& 4.8             &   7.3           &    0.009       \\
NGC 1808& 7.2             &   6.9           &    0.06          \\
NGC 520 & 11.7            &   6.6           &    0.3       \\
NGC 7552& 13.7            &   6.7           &    0.3         \\
NGC 7632& 5.9             &   7.0           &    0.03        \\
NGC 1614& 22.9            &   6.4           &    0.9          \\
NGC 6240& 3.6             &   7.7           &    3.6         \\
Arp 220 & 8.0             &   6.8           &    0.7       \\
\hline
 \end{tabular}
\label{tab:snrate}

\end{table}

\subsection{Qualitative correlation}

\begin{figure*}
\centering
\includegraphics[width=17cm]{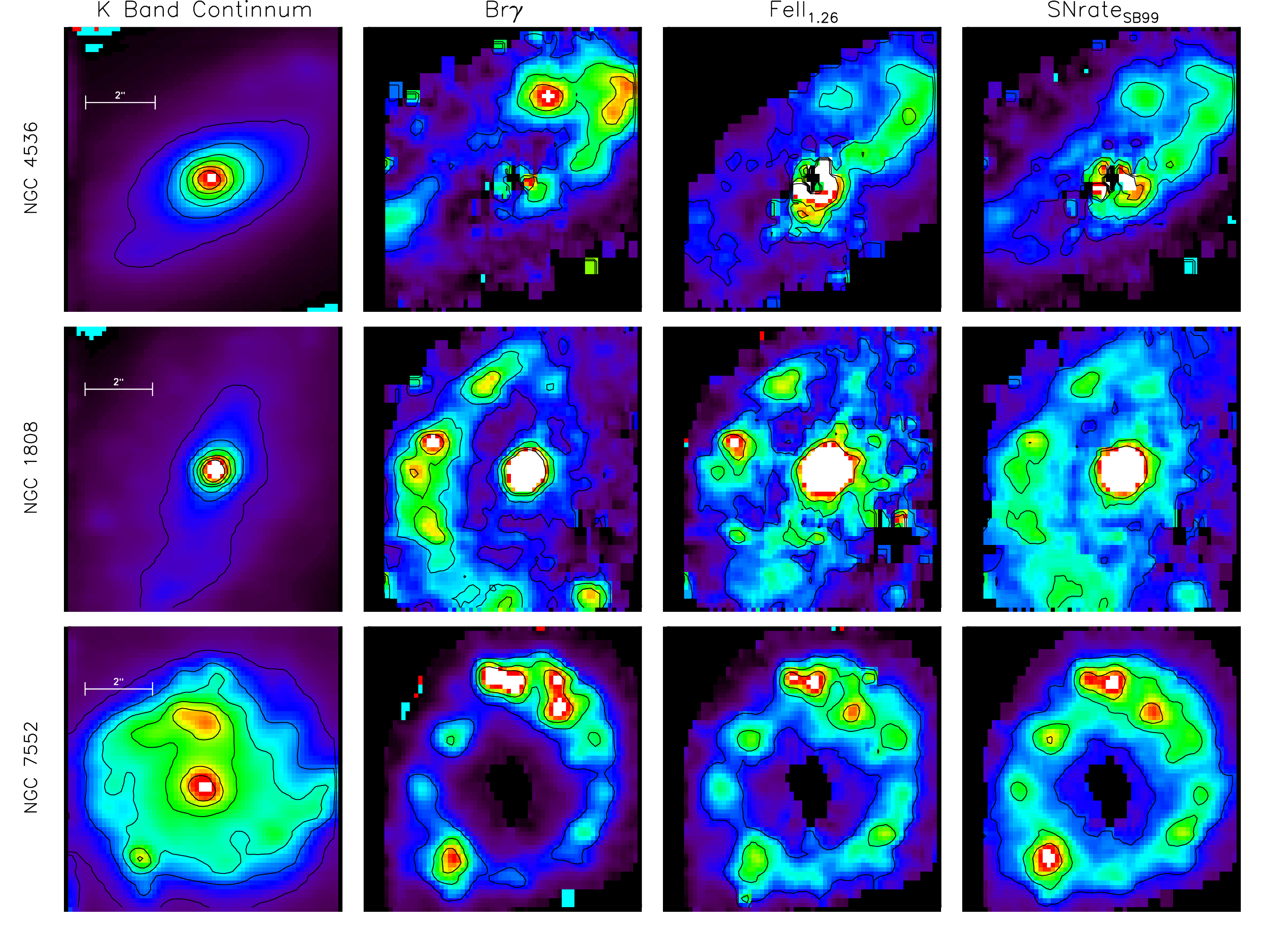}
\caption{A comparison between the K-band continuum, Br$\gamma$
  line map, [FeII]$_{1.26}$ line map, and the SNrate morphologies for
  NGC 4536, NGC 1808, and NGC 7552.  The black discontinuities represent areas that have been filtered due to low signal-to-noise values in the line or extinction maps.  The white pixels represent the highest flux regions.}
\label{fig:snr}
\end{figure*}

To illustrate the morphological relation between the [FeII] emission
and the SNrate, we show a side-by-side comparison of the K-band
continuum, the extinction-corrected Br$\gamma$ flux, the
extinction-corrected [FeII]$_{1.26}$ flux, and the derived SNrate in
Figure~\ref{fig:snr} for the galaxies NGC 4536, NGC 1808, and NGC
7552, which have the highest-quality observations.  A visual
comparison of the morphologies in the four different maps of each
galaxy shows that the SNrate map most closely resembles that of the
[FeII] emission.

For instance, in NGC 4536 there is a bright knot of Br$\gamma$
emission directly north of the nucleus with secondary emission peaks
to the north-west and south-east of the galaxy.  The [FeII] emission
is concentrated on the galaxy nucleus with a long plume of emission
(perhaps an inner spiral arm) extending from the nucleus towards the
north-west.  A small knot of faint [FeII] emission coincides with the
Br$\gamma$ peak.  Mimicking the [FeII] emission, the SNrate peaks at
the center and also shows an arm extending towards the north-west.
The SNrate exhibits a slight increase at the Br$\gamma$ peak, but on
the whole more closely resembles the morphology of [FeII]
emission.

The starburst ring surrounding the bright nucleus of NGC 1808 shows
up clearly in Br$\gamma$.  The ring contains individual bright knots of
Br$\gamma$ emission. The ring is much less prominent in [FeII]
emission.  The SNrate map shows a ring structure more clear than in
the [FeII] map but it is quite diffuse and lacks the contrast that the
Br$\gamma$ knots provide.  The SNrate map also shows diffuse extended
structure around the nucleus, similar to the [FeII] emission but
distinct from the Br$\gamma$ map in which the compact nucleus is much
more isolated.

Finally, the face-on spiral, NGC 7552, also reveals its starburst ring
in Br$\gamma$, with little emission at the nucleus.  In the northern
part of the ring we find three major Br$\gamma$ knots, two slightly
north-west of the nucleus and one elongated knot directly north of the
nucleus.  Comparison shows that only one of the two north-western
Br$\gamma$ knots is discernible in [FeII] emission.  We also see that
the morphology in the elongated knot differs in the two maps.  Focusing on the SNrate morphology, the northern emission knots resemble the relative flux and morphologies of the [FeII] emission much more closely than the Br$\gamma$ emission.  Specifically, the top north-western knot is also missing in SNrate map and the elongated northern Br$\gamma$ peak is resolved to two peaks, resembling the relative flux ratios seen in [FeII].  

All comparisons show that, at least qualitatively, the SNrate maps
resemble the [FeII] maps much more than those of the Br$\gamma$
emission on a pixel-by-pixel basis. Once again we emphasise that the
\emph{only} observational input into the SNrate calculation is the
Br$\gamma$ luminosity and EW(Br$\gamma$); the [FeII] flux is never
used. As the SNrate nevertheless correlates better with the [FeII]
than with the Br$\gamma$ emission, we conclude that this strongly
supports [FeII] as a robust tracer of SNrate.

\subsection{Quantitative correlation}

\begin{figure}
\centering
\resizebox{\hsize}{!}{\includegraphics{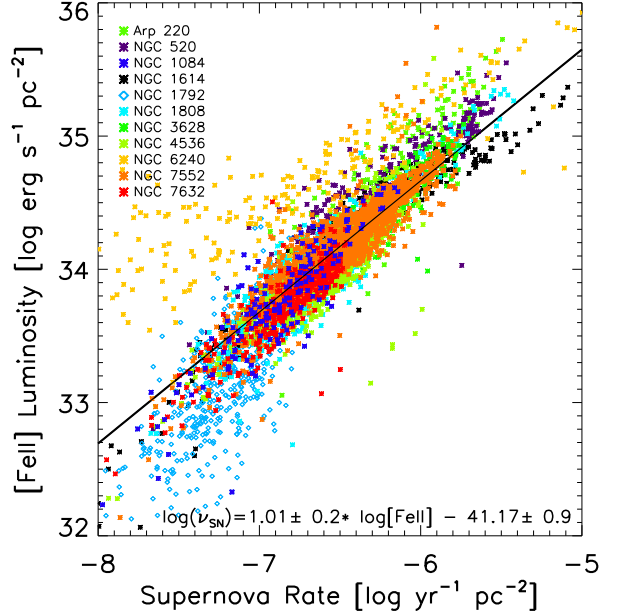}}
\caption{A pixel-pixel plot of SNrate, as derived from SB99, compared
  to [FeII] luminosity.  Each galaxy is represented with a different
  color and the values are normalised to a square parsec.  The black
  line represents the best fit power law excluding NGC 1792, which is
  represented by diamonds.}
\label{fig:pixpix}
\end{figure}

Now that we have established that a qualitative correlation exists
between [FeII] emission and SNrates, it is important to verify to
what extent this relationship holds up quantitatively.  To do this, we
compare in Figure~\ref{fig:pixpix} for each galaxy the [FeII]$_{1.26}$
luminosity and the SNrate derived from Br$\gamma$ pixel-by-pixel.  The
line-maps were not additionally filtered, with the exception of NGC
6240 which was filtered to eliminate pixels where the Br$\gamma$ flux
is insufficient to determine the true extinction.  We normalised both
the SNrate and the [FeII] luminosity to values per square parsec so
that each pixel represents the same physical area.  We performed a
least-squares regression in log space on the combined points from all
sample galaxies except NGC 1792 because of its poor signal-to-noise
data.  The combined linear regression is indicated by a solid black
line in Figure~\ref{fig:pixpix}; it takes the analytical form:

\begin{equation}
log\frac{\nu_{SNrate}}{yr^{-1}pc^{-2}}=\left ( 1.01 \pm 0.2 \right ) \ast
log\frac{[FeII]_{1.26}}{erg s^{-1}pc^{-2}}-41.17 \pm 0.9
\end{equation}

\begin{figure}
\centering
\resizebox{\hsize}{!}{\includegraphics{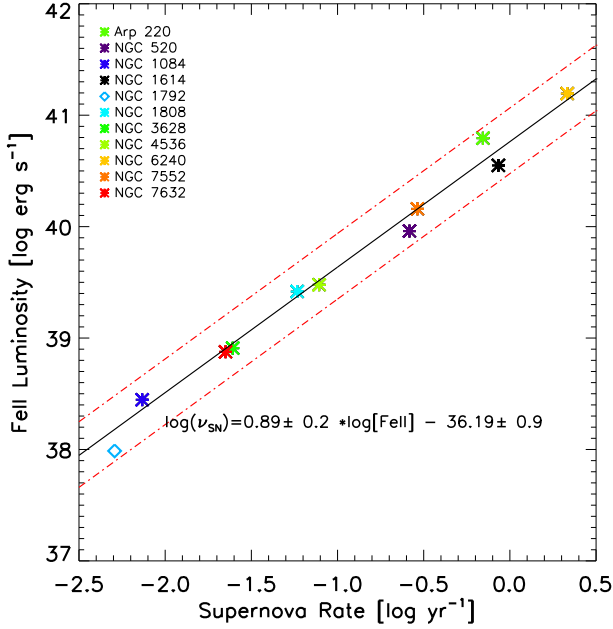}}
\caption{Integrated SNrate plotted against integrated [FeII]
  luminosity over the field of view.  The black line is the
  least-squares regression line, excluding NGC 1792.  The red dotted
  lines represent the least squares regression line if the magnitude
  of extinction was overestimated respectively underestimated by a
  factor of two in optical depth.}
\label{fig:int}
\end{figure}

The errors represent the standard deviation of the slopes and
intercepts of the individual galaxies, thus representing the variation
between individual galaxies' fits.  This is an intrinsic error, corresponding to the variations in properties among the galaxies and not derived from the quality of the observation. To investigate the strength of the relationship on
a global basis, we have plotted , excluding NGC 1792, the integrated
SNrate plotted against the integrated [FeII] luminosity for each
galaxy, again with the best-fit regression line.  Errors in the
extinction are a potentially significant source of uncertainty in the
derived relationship. The dashed lines at either side of this best fit
in Figure~\ref{fig:int} mark the regression lines we would find if the
extinction magnitudes were overestimated respectively underestimated
by factors of two.  It is obvious from the Figure that our result is
essentially unaffected by this uncertainty. The best fit for the global
values is given by:

\begin{equation}
log\frac{\nu_{SNrate}}{yr^{-1}pc^{-2}}=\left ( 0.89 \pm 0.2 \right ) \ast log\frac{[FeII]_{1.26}}{erg s^{-1}pc^{-2}} -36.19 \pm 0.9
\end{equation}

Thus, the averaged and the integrated SNrate - [FeII] luminosity
relation are identical within the errors.  Both procedures yield a
power law with a slope of nearly unity.  The relation is linear
well within the errors.  

An important source of uncertainty in the SNrate calculation is caused
by observational errors in the Br$\gamma$, [FeII] and K band continuum
intensities.  These observational errors are dominated by the
uncertainty in the calibration, which is of the order of 10\% of the
flux in all cases except NGC 1792, where the errors are dominated by
noise, estimated to be about 20\% of the flux.  The uncertainties in
the derived SNrate are the same as those in Br$\gamma$ flux, i.e.
about $10\%$.  However, the observational errors are not the dominant
source of uncertainty. This is the systematic error introduced by the
choice of the starburst model.  Because we use SB99, our choices are
limited to two star-formation models only, instantaneous and
continuous.  In the next section, we discuss the effects of the choice
of the burst model on the relation between SNrate and [FeII] luminosity.

\section{Dependency on burst model}
\label{sec:cont}
\begin{figure}
\centering
\resizebox{\hsize}{!}{\includegraphics{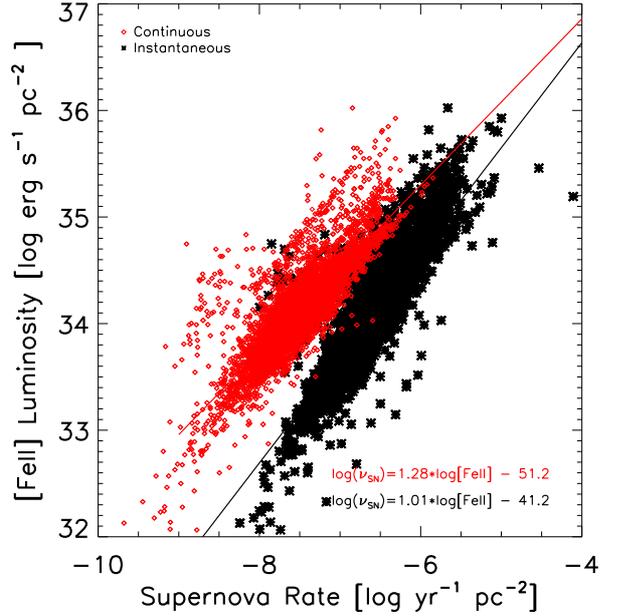}}
\caption{Pixel-pixel comparison of SNrate to [FeII] luminosity.  The
  black points represent the pixel values for all the galaxies
  using the instantaneous-burst model, while the red points represent
  the same pixels using the continuous-burst model.}
\label{fig:cont}
\end{figure}

So far, we have based our analysis on the assumption that star
formation in the sample galaxies is best represented by an
instantaneous burst.  The other extreme case also modelled by SB99 is
the continuous star-formation scenario.  It is unlikely that either of
these extreme cases is a true representation of the situation in the
galaxies considered; we suspect that reality is somewhere in between
and better described by one or more star bursts extended in time.  We
cannot determine the actual star formation history of the sample, but
the SB99 model allows us to constrain the validity of our
determination of SNrate/[FeII] luminosity relation by performing the
same analysis this time assuming a continuous star-formation mode.
The result is shown in (Figure~\ref{fig:cont}) as a plot of the [FeII]
luminosity versus the SNrate for the combined pixels of all sample
galaxies. In this Figure, the black dots represent the SNrates
assuming an instantaneous burst (as in Figure~\ref{fig:pixpix}) and
the red points represent the same pixels where the SNrate is
calculated assuming continuous star formation.

From Figure~\ref{fig:cont} it is clear that also in continuous
star-formation mode, the SNrate is still closely related to the [FeII]
luminosity.  However, in this case, the relation is no longer linear
as the best fit slope is 1.20. We also note that in this situation, a
given [FeII] luminosity also implies a SNrate roughly an order of
magnitude higher than in the instantaneous burst model.

\section{Comparison to Radio SNrate}
\label{sec:radio}
\begin{figure}
\centering
\resizebox{\hsize}{!}{\includegraphics{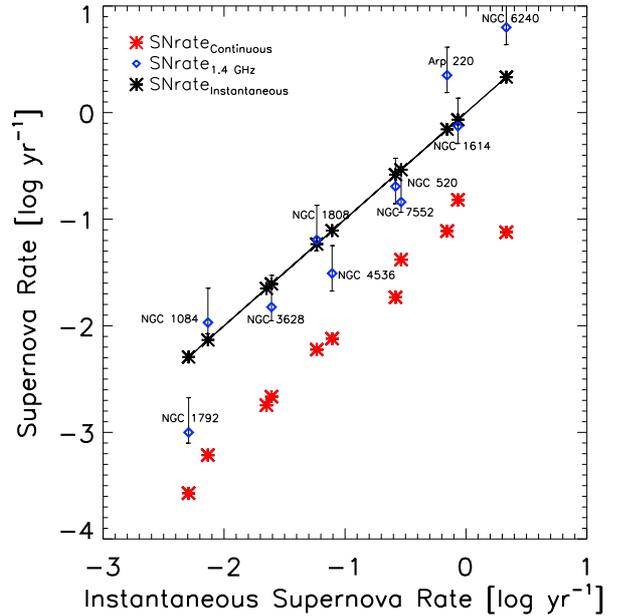}}
\caption{Comparison of the SB99 instantaneous SNrate (x-axis) to the
  radio SNrate (blue diamond), the SB99 continuous SNrate (red
  asterisk) and for comparison the SB99 instantaneous SNrate (black
  asterisk).}
\label{fig:radio1}
\end{figure}

Which mode best represents our sample? Radio continuum emission is the
classical tracer of SNrate, and radio flux densities can be used to
obtain an independent estimate of the SNrate for comparison to the
SNrates derived with the SB99 models.  We apply the equation given by
\citet{1994ApJ...424..114H}, which relates non-thermal radio
luminosity and SNrate, to VLA observations.  There are no VLA archived
observation of NGC 7632 and this galaxy is excluded from the
comparison.  A similar situation applies to NGC 7552, but here we
could use the ATCA 4.8 GHz map instead.  For the nearest extended
galaxies, we matched the radio integration area to that of SINFONI.
For the more distant NGC 1614, NGC 6240, and Arp 220, we used NED VLA
1.4 GHz integrated flux densities since these galaxies are compact
enough that the entire galaxy is encompassed in the SINFONI field of
view.  In Figure~\ref{fig:radio1} we compare the radio SNrates thus
derived to the SB99 instantaneous and continuous SNrates.

For all galaxies, except NGC 1792, NGC 4536, NGC 6240 and Arp 220, the
radio SNrate closely matches those given by the SB99
instantaneous-burst model.  As noted before, the NIR-emission line-maps
of NGC 1792 are too noisy to be reliable.  However, NGC 4536 is a
nearly face-on galaxy with relatively low extinction and high
signal-to-noise spectra, and the NIR data should be reliable. Thus,
NGC 4536 is undergoing star formation that is closer to instantaneous but extended in time.  Arp 220 and NGC 6240 have radio
SNrates even higher than the SB99 instantaneous SNrates.  Also, in Figure~\ref{fig:pixpix}, the points for these galaxies lie systematically above the best fit line, with a similar slope to the rest of the galaxies.  This offset can be explained by either an excess of [FeII] emission or a deficiency of Br$\gamma$ emission, since these are the two observables that are used to scale SNrate and [FeII] luminosity. In the case of Br$\gamma$ deficiency, the Br$\gamma$ emission could be selectively suppressed by dust absorption of Lyman continuum photons in these very dense systems.  However, in NGC 6240 and Arp 220, it is tempting to speculate that this excess 
is caused by merger-related shocks that are over-exciting the [FeII] in addition to the SNRs.  Evidence for large-scale shocks in NGC 6240 comes from the prominent wings on the H$_2$ 1-0 S(1) line which trace the shocked superwind \citep{1993ApJ...405..522V,2010A&A...524A..56E}.  The [FeII] emission is morphologically similar to the H$_2$, indicating that the [FeII] also traces the superwind.  These powerful merger (U)LIRGs may demonstrate a limit to the
validity of the direct relationship between SNrate and [FeII] luminosity.  In addition, determining an accurate extinction in (U)LIRGs presents a challenge.  The under or over estimation of extinction leads to inaccurate [FeII] luminosities as well as Br$\gamma$ luminosities used to scale the SNrate.  
However, the offset between the (U)LIRGs and the other galaxies is minor.  They share a common slope and appear to be offset only in [FeII] emission.  The validity of this relationship in the case of (U)LIRGs can only be established
by studying a larger sample of these galaxies.  In any case, it appears that
the majority of (modest) starburst galaxies is well-represented by
the assumption of a (nearly) instantaneous burst of star formation.
In addition, the very good agreement between SNrates derived from the
radio continuum, and from NIR data provides added confidence in the
diagnostic strength of [FeII] as a tracer of SNrate.

\section{Conclusion}
\label{sec:conclusion}
Using SINFONI observations of 11 near-by galaxies, we have performed a
pixel-by-pixel analysis of the correlation between [FeII]$_{1.26}$
emission and independently derived SNrates.  We measured accurate
Br$\gamma$, Pa$\beta$, [FeII]$_{1.26}$, [FeII]$_{1.64}$, and
H$_{2,2.12}$ line fluxes.  We determined Br$\gamma$ equivalent widths
which were used as input into the SB99 model to find the starburst
age in each pixel.  In the SB99 model, this defines a SNrate,
normalised to an initial mass of $10^6 M_{\odot}$, which we scaled
with the observed Br$\gamma$ luminosity to find the actual SNrate in
each pixel.

The comparison of the [FeII]$_{1.26}$ luminosity to the SNrate (derived
only from Br$\gamma$ equivalent width and luminosity) reveals a nearly
linear correlation, where the error represents the spread of intrinsic properties over our sample of galaxies.

\[ log\frac{\nu_{SNrate}}{yr^{-1}pc^{-2}} = 1.01 \pm 0.2\ast  log\frac{[FeII]_{1.26}}{erg s^{-1}pc^{-2}} - 41.17 \pm 0.9\] 
This relationship is valid both on a pixel-pixel basis and for the
integrated galaxy.  For the integrated [FeII] luminosity and SNrate,
the fit is remarkably tight with very little spread.  However, to
correctly determine the absolute SNrates, it is still critical to
determine whether star formation has occurred in a (nearly)
instantaneous burst, or has proceeded in a continuous fashion.
SNrates derived from radio continuum observations may be used to
distinguish these scenarios.  Most of the modest starburst galaxies
in our sample are best fitted assuming instantaneous star formation.  However, we find that the relationship breaks down for (U)LIRGs in our sample.

From the strong correlations found in this sample of galaxies, we
confidently conclude that [FeII]$_{1.26}$ emission line strengths are
quantitatively correlated with SNrates and can be used to derive these
rates either locally or globally.  This use of [FeII] as a robust
tracer of SNrate provides us with a very useful diagnostic tool.  It
is particularly important as it allows determination of SNrates from
NIR observations of distant galaxies, where individual SNRs can not be
resolved.

\bibliographystyle{aa}
\bibliography{bib_file.bib}

\begin{thebibliography}{76}
\expandafter\ifx\csname natexlab\endcsname\relax\def\natexlab#1{#1}\fi

\bibitem[{{Alonso-Herrero} {et~al.}(2001){Alonso-Herrero}, {Engelbracht},
  {Rieke}, {Rieke}, \& {Quillen}}]{2001ApJ...546..952A}
{Alonso-Herrero}, A., {Engelbracht}, C.~W., {Rieke}, M.~J., {Rieke}, G.~H., \&
  {Quillen}, A.~C. 2001, \apj, 546, 952

\bibitem[{{Alonso-Herrero} {et~al.}(2003){Alonso-Herrero}, {Rieke}, {Rieke}, \&
  {Kelly}}]{2003AJ....125.1210A}
{Alonso-Herrero}, A., {Rieke}, G.~H., {Rieke}, M.~J., \& {Kelly}, D.~M. 2003,
  \aj, 125, 1210

\bibitem[{{Alonso-Herrero} {et~al.}(2000){Alonso-Herrero}, {Rieke}, {Rieke}, \&
  {Scoville}}]{2000ApJ...532..845A}
{Alonso-Herrero}, A., {Rieke}, G.~H., {Rieke}, M.~J., \& {Scoville}, N.~Z.
  2000, \apj, 532, 845

\bibitem[{{Arp}(1981)}]{1981ApJS...46...75A}
{Arp}, H. 1981, \apjs, 46, 75

\bibitem[{Awaki \& Koyama(1993)}]{Awaki1993221}
Awaki, H. \& Koyama, K. 1993, Advances in Space Research, 13, 221

\bibitem[{{Beswick} {et~al.}(2001){Beswick}, {Pedlar}, {Mundell}, \&
  {Gallimore}}]{2001MNRAS.325..151B}
{Beswick}, R.~J., {Pedlar}, A., {Mundell}, C.~G., \& {Gallimore}, J.~F. 2001,
  \mnras, 325, 151

\bibitem[{{Calzetti}(1997)}]{1997AJ....113..162C}
{Calzetti}, D. 1997, \aj, 113, 162

\bibitem[{{Carral} {et~al.}(1990){Carral}, {Turner}, \&
  {Ho}}]{1990ApJ...362..434C}
{Carral}, P., {Turner}, J.~L., \& {Ho}, P.~T.~P. 1990, \apj, 362, 434

\bibitem[{{Condon}(1987)}]{1987ApJS...65..485C}
{Condon}, J.~J. 1987, \apjs, 65, 485

\bibitem[{{Condon}(1992)}]{1992ARA&A..30..575C}
{Condon}, J.~J. 1992, \araa, 30, 575

\bibitem[{{Condon} {et~al.}(1982){Condon}, {Condon}, {Gisler}, \&
  {Puschell}}]{1982ApJ...252..102C}
{Condon}, J.~J., {Condon}, M.~A., {Gisler}, G., \& {Puschell}, J.~J. 1982,
  \apj, 252, 102

\bibitem[{{Condon} \& {Yin}(1990)}]{1990ApJ...357...97C}
{Condon}, J.~J. \& {Yin}, Q.~F. 1990, \apj, 357, 97

\bibitem[{{Corwin} {et~al.}(1994){Corwin}, {Buta}, \& {de Vaucouleurs}}]{morph}
{Corwin}, Jr., H.~G., {Buta}, R.~J., \& {de Vaucouleurs}, G. 1994, \aj, 108,
  2128

\bibitem[{{Dahlem} {et~al.}(1994){Dahlem}, {Bomans}, \&
  {Will}}]{1994ApJ...432..590D}
{Dahlem}, M., {Bomans}, D.~J., \& {Will}, J. 1994, \apj, 432, 590

\bibitem[{{Downes} \& {Solomon}(1998)}]{1998ApJ...507..615D}
{Downes}, D. \& {Solomon}, P.~M. 1998, \apj, 507, 615

\bibitem[{{Doyon} {et~al.}(1994){Doyon}, {Wells}, {Wright}, {Joseph}, {Nadeau},
  \& {James}}]{1994ApJ...437L..23D}
{Doyon}, R., {Wells}, M., {Wright}, G.~S., {et~al.} 1994, \apjl, 437, L23

\bibitem[{{Eales} {et~al.}(1990){Eales}, {Becklin}, {Hodapp}, {Simons}, \&
  {Wynn-Williams}}]{1990ApJ...365..478E}
{Eales}, S.~A., {Becklin}, E.~E., {Hodapp}, K.-W., {Simons}, D.~A., \&
  {Wynn-Williams}, C.~G. 1990, \apj, 365, 478

\bibitem[{{Emerson} {et~al.}(1984){Emerson}, {Clegg}, {Gee}, {Griffin},
  {Cunningham}, {Brown}, {Robson}, \& {Longmore}}]{1984Natur.311..237E}
{Emerson}, J.~P., {Clegg}, P.~E., {Gee}, G., {et~al.} 1984, \nat, 311, 237

\bibitem[{{Engel} {et~al.}(2010){Engel}, {Davies}, {Genzel}, {Tacconi},
  {Hicks}, {Sturm}, {Naab}, {Johansson}, {Karl}, {Max}, {Medling}, \& {van der
  Werf}}]{2010A&A...524A..56E}
{Engel}, H., {Davies}, R.~I., {Genzel}, R., {et~al.} 2010, \aap, 524, A56

\bibitem[{{Eskridge} {et~al.}(2002){Eskridge}, {Frogel}, {Pogge}, {Quillen},
  {Berlind}, {Davies}, {DePoy}, {Gilbert}, {Houdashelt}, {Kuchinski},
  {Ram{\'{\i}}rez}, {Sellgren}, {Stutz}, {Terndrup}, \&
  {Tiede}}]{2002ApJS..143...73E}
{Eskridge}, P.~B., {Frogel}, J.~A., {Pogge}, R.~W., {et~al.} 2002, \apjs, 143,
  73

\bibitem[{{Feinstein} {et~al.}(1990){Feinstein}, {Mendez}, {Vega}, \&
  {Forte}}]{1990A&A...239...90F}
{Feinstein}, C., {Mendez}, M., {Vega}, I., \& {Forte}, J.~C. 1990, \aap, 239,
  90

\bibitem[{{Forbes} {et~al.}(1992){Forbes}, {Boisson}, \&
  {Ward}}]{1992MNRAS.259..293F}
{Forbes}, D.~A., {Boisson}, C., \& {Ward}, M.~J. 1992, \mnras, 259, 293

\bibitem[{{Forbes} {et~al.}(1994{\natexlab{a}}){Forbes}, {Kotilainen}, \&
  {Moorwood}}]{1994ApJ...433L..13F}
{Forbes}, D.~A., {Kotilainen}, J.~K., \& {Moorwood}, A.~F.~M.
  1994{\natexlab{a}}, \apjl, 433, L13

\bibitem[{{Forbes} {et~al.}(1994{\natexlab{b}}){Forbes}, {Norris}, {Williger},
  \& {Smith}}]{1994AJ....107..984F}
{Forbes}, D.~A., {Norris}, R.~P., {Williger}, G.~M., \& {Smith}, R.~C.
  1994{\natexlab{b}}, \aj, 107, 984

\bibitem[{{Forbes} {et~al.}(1993){Forbes}, {Ward}, {Rotaciuc}, {Blietz},
  {Genzel}, {Drapatz}, {Van der Werf}, \& {Krabbe}}]{1993ApJ...406L..11F}
{Forbes}, D.~A., {Ward}, M.~J., {Rotaciuc}, V., {et~al.} 1993, \apjl, 406, L11

\bibitem[{{Genzel} {et~al.}(1998){Genzel}, {Lutz}, {Sturm}, {Egami}, {Kunze},
  {Moorwood}, {Rigopoulou}, {Spoon}, {Sternberg}, {Tacconi-Garman}, {Tacconi},
  \& {Thatte}}]{1998ApJ...498..579G}
{Genzel}, R., {Lutz}, D., {Sturm}, E., {et~al.} 1998, \apj, 498, 579

\bibitem[{{Gonz{\'a}lez-Mart{\'{\i}}n}
  {et~al.}(2006){Gonz{\'a}lez-Mart{\'{\i}}n}, {Masegosa}, {M{\'a}rquez},
  {Guerrero}, \& {Dultzin-Hacyan}}]{2006A&A...460...45G}
{Gonz{\'a}lez-Mart{\'{\i}}n}, O., {Masegosa}, J., {M{\'a}rquez}, I.,
  {Guerrero}, M.~A., \& {Dultzin-Hacyan}, D. 2006, \aap, 460, 45

\bibitem[{{Graham} {et~al.}(1987){Graham}, {Wright}, \&
  {Longmore}}]{1987ApJ...313..847G}
{Graham}, J.~R., {Wright}, G.~S., \& {Longmore}, A.~J. 1987, \apj, 313, 847

\bibitem[{{Graham} {et~al.}(1990){Graham}, {Wright}, \&
  {Longmore}}]{1990ApJ...352..172G}
{Graham}, J.~R., {Wright}, G.~S., \& {Longmore}, A.~J. 1990, \apj, 352, 172

\bibitem[{{Greenhouse} {et~al.}(1997){Greenhouse}, {Satyapal}, {Woodward},
  {Fischer}, {Thompson}, {Forrest}, {Pipher}, {Raines}, {Smith}, {Watson}, \&
  {Rudy}}]{1997ApJ...485..438G}
{Greenhouse}, M.~A., {Satyapal}, S., {Woodward}, C.~E., {et~al.} 1997, \apj,
  485, 438

\bibitem[{{Greenhouse} {et~al.}(1991){Greenhouse}, {Woodward}, {Thronson},
  {Rudy}, {Rossano}, {Erwin}, \& {Puetter}}]{1991ApJ...383..164G}
{Greenhouse}, M.~A., {Woodward}, C.~E., {Thronson}, Jr., H.~A., {et~al.} 1991,
  \apj, 383, 164

\bibitem[{{Haynes} {et~al.}(1979){Haynes}, {Giovanelli}, \&
  {Roberts}}]{1979ApJ...229...83H}
{Haynes}, M.~P., {Giovanelli}, R., \& {Roberts}, M.~S. 1979, \apj, 229, 83

\bibitem[{{Huang} {et~al.}(1994){Huang}, {Thuan}, {Chevalier}, {Condon}, \&
  {Yin}}]{1994ApJ...424..114H}
{Huang}, Z.~P., {Thuan}, T.~X., {Chevalier}, R.~A., {Condon}, J.~J., \& {Yin},
  Q.~F. 1994, \apj, 424, 114

\bibitem[{{Hughes} {et~al.}(2005){Hughes}, {Axon}, {Atkinson},
  {Alonso-Herrero}, {Scarlata}, {Marconi}, {Batcheldor}, {Binney}, {Capetti},
  {Carollo}, {Dressel}, {Gerssen}, {Macchetto}, {Maciejewski}, {Merrifield},
  {Ruiz}, {Sparks}, {Stiavelli}, \& {Tsvetanov}}]{2005AJ....130...73H}
{Hughes}, M.~A., {Axon}, D., {Atkinson}, J., {et~al.} 2005, \aj, 130, 73

\bibitem[{{Hummel}(1980)}]{1980A&AS...41..151H}
{Hummel}, E. 1980, \aaps, 41, 151

\bibitem[{{Hummer} \& {Storey}(1987)}]{1987MNRAS.224..801H}
{Hummer}, D.~G. \& {Storey}, P.~J. 1987, \mnras, 224, 801

\bibitem[{{Israel}(2009)}]{2009A&A...506..689I}
{Israel}, F.~P. 2009, \aap, 506, 689

\bibitem[{{Kotilainen} {et~al.}(1996){Kotilainen}, {Forbes}, {Moorwood}, {Van
  der Werf}, \& {Ward}}]{1996A&A...313..771K}
{Kotilainen}, J.~K., {Forbes}, D.~A., {Moorwood}, A.~F.~M., {Van der Werf},
  P.~P., \& {Ward}, M.~J. 1996, \aap, 313, 771

\bibitem[{{Kotilainen} {et~al.}(2001){Kotilainen}, {Reunanen}, {Laine}, \&
  {Ryder}}]{2001A&A...366..439K}
{Kotilainen}, J.~K., {Reunanen}, J., {Laine}, S., \& {Ryder}, S.~D. 2001, \aap,
  366, 439

\bibitem[{{Krabbe} {et~al.}(1994){Krabbe}, {Sternberg}, \&
  {Genzel}}]{1994ApJ...425...72K}
{Krabbe}, A., {Sternberg}, A., \& {Genzel}, R. 1994, \apj, 425, 72

\bibitem[{{Laine} {et~al.}(2006){Laine}, {Kotilainen}, {Reunanen}, {Ryder}, \&
  {Beck}}]{2006AJ....131..701L}
{Laine}, S., {Kotilainen}, J.~K., {Reunanen}, J., {Ryder}, S.~D., \& {Beck}, R.
  2006, \aj, 131, 701

\bibitem[{{Leitherer} {et~al.}(1999){Leitherer}, {Schaerer}, {Goldader},
  {Gonz{\'a}lez Delgado}, {Robert}, {Kune}, {de Mello}, {Devost}, \&
  {Heckman}}]{sb99}
{Leitherer}, C., {Schaerer}, D., {Goldader}, J.~D., {et~al.} 1999, \apjs, 123,
  3

\bibitem[{{Liu} \& {Bregman}(2005)}]{2005ApJS..157...59L}
{Liu}, J. \& {Bregman}, J.~N. 2005, \apjs, 157, 59

\bibitem[{{Martin} \& {Whittet}(1990)}]{1990ApJ...357..113M}
{Martin}, P.~G. \& {Whittet}, D.~C.~B. 1990, \apj, 357, 113

\bibitem[{{Moorwood} \& {Oliva}(1988)}]{1988A&A...203..278M}
{Moorwood}, A.~F.~M. \& {Oliva}, E. 1988, \aap, 203, 278

\bibitem[{{Moorwood} {et~al.}(1988){Moorwood}, {Oliva}, \&
  {Danziger}}]{1988srim.conf..391M}
{Moorwood}, A.~F.~M., {Oliva}, E., \& {Danziger}, I.~J. 1988, in IAU Colloq.
  101: Supernova Remnants and the Interstellar Medium, ed. {R.~S.~Roger \&
  T.~L.~Landecker}, 391--+

\bibitem[{{Mouri} {et~al.}(2000){Mouri}, {Kawara}, \&
  {Taniguchi}}]{2000ApJ...528..186M}
{Mouri}, H., {Kawara}, K., \& {Taniguchi}, Y. 2000, \apj, 528, 186

\bibitem[{{Neff} {et~al.}(1990){Neff}, {Hutchings}, {Standord}, \&
  {Unger}}]{1990AJ.....99.1088N}
{Neff}, S.~G., {Hutchings}, J.~B., {Standord}, S.~A., \& {Unger}, S.~W. 1990,
  \aj, 99, 1088

\bibitem[{{Norman} {et~al.}(1996){Norman}, {Bowen}, {Heckman}, {Blades}, \&
  {Danly}}]{1996ApJ...472...73N}
{Norman}, C.~A., {Bowen}, D.~V., {Heckman}, T., {Blades}, C., \& {Danly}, L.
  1996, \apj, 472, 73

\bibitem[{{Nussbaumer} \& {Storey}(1988)}]{1988A&A...193..327N}
{Nussbaumer}, H. \& {Storey}, P.~J. 1988, \aap, 193, 327

\bibitem[{{Oliva} {et~al.}(1989){Oliva}, {Moorwood}, \&
  {Danziger}}]{1989A&A...214..307O}
{Oliva}, E., {Moorwood}, A.~F.~M., \& {Danziger}, I.~J. 1989, \aap, 214, 307

\bibitem[{{Oliva} {et~al.}(1990){Oliva}, {Moorwood}, \&
  {Danziger}}]{1990A&A...240..453O}
{Oliva}, E., {Moorwood}, A.~F.~M., \& {Danziger}, I.~J. 1990, \aap, 240, 453

\bibitem[{{Olsson} {et~al.}(2010){Olsson}, {Aalto}, {Thomasson}, \&
  {Beswick}}]{2010A&A...513A..11O}
{Olsson}, E., {Aalto}, S., {Thomasson}, M., \& {Beswick}, R. 2010, \aap, 513,
  A11+

\bibitem[{{Phillips}(1993)}]{1993AJ....105..486P}
{Phillips}, A.~C. 1993, \aj, 105, 486

\bibitem[{{Prugniel} {et~al.}(1998){Prugniel}, {Zasov}, {Busarello}, \&
  {Simien}}]{LEDA}
{Prugniel}, P., {Zasov}, A., {Busarello}, G., \& {Simien}, F. 1998, \aaps, 127,
  117

\bibitem[{{Puxley} {et~al.}(1988){Puxley}, {Hawarden}, \&
  {Mountain}}]{1988MNRAS.234P..29P}
{Puxley}, P.~J., {Hawarden}, T.~G., \& {Mountain}, C.~M. 1988, \mnras, 234, 29P

\bibitem[{{Ramya} {et~al.}(2007){Ramya}, {Sahu}, \&
  {Prabhu}}]{2007MNRAS.381..511R}
{Ramya}, S., {Sahu}, D.~K., \& {Prabhu}, T.~P. 2007, \mnras, 381, 511

\bibitem[{{Rangwala} {et~al.}(2011){Rangwala}, {Maloney}, {Glenn}, {Wilson},
  {Rykala}, {Isaak}, {Baes}, {Bendo}, {Boselli}, {Bradford}, {Clements},
  {Cooray}, {Fulton}, {Imhof}, {Kamenetzky}, {Madden}, {Mentuch}, {Sacchi},
  {Sauvage}, {Schirm}, {Smith}, {Spinoglio}, \&
  {Wolfire}}]{2011ApJ...743...94R}
{Rangwala}, N., {Maloney}, P.~R., {Glenn}, J., {et~al.} 2011, \apj, 743, 94

\bibitem[{{Reuter} {et~al.}(1991){Reuter}, {Krause}, {Wielebinski}, \&
  {Lesch}}]{1991A&A...248...12R}
{Reuter}, H., {Krause}, M., {Wielebinski}, R., \& {Lesch}, H. 1991, \aap, 248,
  12

\bibitem[{{Roberts} {et~al.}(2001){Roberts}, {Schurch}, \&
  {Warwick}}]{2001MNRAS.324..737R}
{Roberts}, T.~P., {Schurch}, N.~J., \& {Warwick}, R.~S. 2001, \mnras, 324, 737

\bibitem[{{Rots}(1978)}]{1978AJ.....83..219R}
{Rots}, A.~H. 1978, \aj, 83, 219

\bibitem[{{Sanders} {et~al.}(2003){Sanders}, {Mazzarella}, {Kim}, {Surace}, \&
  {Soifer}}]{2003AJ....126.1607S}
{Sanders}, D.~B., {Mazzarella}, J.~M., {Kim}, D.-C., {Surace}, J.~A., \&
  {Soifer}, B.~T. 2003, \aj, 126, 1607

\bibitem[{{Schinnerer} {et~al.}(1997){Schinnerer}, {Eckart}, {Quirrenbach},
  {Boker}, {Tacconi-Garman}, {Krabbe}, \& {Sternberg}}]{1997ApJ...488..174S}
{Schinnerer}, E., {Eckart}, A., {Quirrenbach}, A., {et~al.} 1997, \apj, 488,
  174

\bibitem[{{Schulz} {et~al.}(1993){Schulz}, {Fried}, {R{\"o}ser}, \&
  {Keel}}]{1993A&A...277..416S}
{Schulz}, H., {Fried}, J.~W., {R{\"o}ser}, S., \& {Keel}, W.~C. 1993, \aap,
  277, 416

\bibitem[{{Scoville} {et~al.}(1997){Scoville}, {Yun}, \&
  {Bryant}}]{1997ApJ...484..702S}
{Scoville}, N.~Z., {Yun}, M.~S., \& {Bryant}, P.~M. 1997, \apj, 484, 702

\bibitem[{{Shull} \& {Draine}(1987)}]{1987ASSL..134..283S}
{Shull}, J.~M. \& {Draine}, B.~T. 1987, in Astrophysics and Space Science
  Library, Vol. 134, Interstellar Processes, ed. {D.~J.~Hollenbach \&
  H.~A.~Thronson Jr.}, 283--319

\bibitem[{{Smith} {et~al.}(1998){Smith}, {Lonsdale}, {Lonsdale}, \&
  {Diamond}}]{1998ApJ...493L..17S}
{Smith}, H.~E., {Lonsdale}, C.~J., {Lonsdale}, C.~J., \& {Diamond}, P.~J. 1998,
  \apjl, 493, L17

\bibitem[{{Soifer} {et~al.}(1984){Soifer}, {Neugebauer}, {Helou}, {Lonsdale},
  {Hacking}, {Rice}, {Houck}, {Low}, \& {Rowan-Robinson}}]{1984ApJ...283L...1S}
{Soifer}, B.~T., {Neugebauer}, G., {Helou}, G., {et~al.} 1984, \apjl, 283, L1

\bibitem[{{Stanford} \& {Balcells}(1990)}]{1990ApJ...355...59S}
{Stanford}, S.~A. \& {Balcells}, M. 1990, \apj, 355, 59

\bibitem[{{Stanford} \& {Balcells}(1991)}]{1991ApJ...370..118S}
{Stanford}, S.~A. \& {Balcells}, M. 1991, \apj, 370, 118

\bibitem[{{Sturm} {et~al.}(1996){Sturm}, {Lutz}, {Genzel}, {Sternberg},
  {Egami}, {Kunze}, {Rigopoulou}, {Bauer}, {Feuchtgruber}, {Moorwood}, \& {de
  Graauw}}]{1996A&A...315L.133S}
{Sturm}, E., {Lutz}, D., {Genzel}, R., {et~al.} 1996, \aap, 315, L133

\bibitem[{{Van der Werf} {et~al.}(1993){Van der Werf}, {Genzel}, {Krabbe},
  {Blietz}, {Lutz}, {Drapatz}, {Ward}, \& {Forbes}}]{1993ApJ...405..522V}
{Van der Werf}, P.~P., {Genzel}, R., {Krabbe}, A., {et~al.} 1993, \apj, 405,
  522

\bibitem[{{Vanzi} \& {Rieke}(1997)}]{1997ApJ...479..694V}
{Vanzi}, L. \& {Rieke}, G.~H. 1997, \apj, 479, 694

\bibitem[{{Veron-Cetty} \& {Veron}(1985)}]{1985A&A...145..425V}
{Veron-Cetty}, M. \& {Veron}, P. 1985, \aap, 145, 425

\bibitem[{{Vila} {et~al.}(1990){Vila}, {Pedlar}, {Davies}, {Hummel}, \&
  {Axon}}]{1990MNRAS.242..379V}
{Vila}, M.~B., {Pedlar}, A., {Davies}, R.~D., {Hummel}, E., \& {Axon}, D.~J.
  1990, \mnras, 242, 379

\bibitem[{{Yuan} {et~al.}(2010){Yuan}, {Kewley}, \&
  {Sanders}}]{2010ApJ...709..884Y}
{Yuan}, T.-T., {Kewley}, L.~J., \& {Sanders}, D.~B. 2010, \apj, 709, 884

\end{thebibliography}
\listofobjects

\end{document}